\documentclass[11pt,a4paper]{article}

\pdfoutput=1
\usepackage{jheppub}
\usepackage{amsfonts}
\usepackage{amsmath}
\usepackage{amssymb}
\usepackage{graphicx,color}
\usepackage{float}
\usepackage{hyperref}
\usepackage{subfigure}

\usepackage{slashed}
\usepackage{mathtools}
\usepackage{hhline}
\usepackage{xcolor}
\usepackage{ulem}

\newcommand{\bq}{\begin{eqnarray}}
\newcommand{\nq}{\end{eqnarray}}
\newcommand{\loopk}{\int \frac{d^4k}{(2\pi)^4}}
\newcommand{\loopp}{\int \frac{d^4p}{(2\pi)^4}}

\newcommand{\looppf}{\int \frac{d^4p}{(2\pi)^4}}
\newcommand{\looppt}{\int \frac{d^3p}{(2\pi)^3}}
\newcommand{\fproppone}{\frac{i(\slashed{p}-M_1)}{p^2-M_1^2}}
\newcommand{\fpropptwo}{\frac{i(\slashed{p}-M_2)}{p^2-M_2^2}}

\title{Warm Inflation, Neutrinos and Dark matter: \\ a minimal extension of the Standard Model}

\author[a]{Miguel Levy}    
\author[b,c]{Jo\~{a}o G.~Rosa}    
\author[c,d,e]{Lu\'{\i}s B. Ventura}

\affiliation[a]{Centro de F\'isica Te\'orica de Part\'iculas-CFTP and Departamento de
  F\'isica,  Instituto Superior T\'ecnico, Universidade de Lisboa, Av  Rovisco Pais, 1, P-1049-001 Lisboa, Portugal}
\affiliation[b]{Univ Coimbra, Faculdade de Ci\^encias e Tecnologia da Universidade de Coimbra, Rua Larga, 3004-516 Coimbra, Portuga}
\affiliation[c]{CFisUC, Rua Larga, 3004-516 Coimbra, Portugal}
\affiliation[d]{Departamento de F\'{\i}sica da Universidade de Aveiro,  Campus de Santiago, 3810-183 Aveiro, Portugal}
\affiliation[e]{CIDMA,  Campus de Santiago, 3810-183 Aveiro, Portugal}

\emailAdd{miguelplevy@ist.utl.pt}
\emailAdd{jgrosa@uc.pt}
\emailAdd{lbventura@ua.pt}

\abstract{
	We show that warm inflation can be realized within a minimal extension of the Standard Model with three right-handed neutrinos, three complex scalars and a gauged lepton/B-L U(1) symmetry. This simple model can address all the shortcomings of the Standard Model that are not related to fine-tuning, within general relativity, with distinctive experimental signatures that can be probed in the near future. The inflaton field emerges from the collective breaking of the U(1) symmetry, and interacts with two of the right-handed neutrinos, sustaining a high-temperature radiation bath during inflation. The discrete interchange symmetry of the model protects the scalar potential against large thermal corrections and leads to a stable inflaton remnant at late times which can account for dark matter. Consistency of the model and agreement with Cosmic Microwave Background observations naturally yield light neutrino masses below 0.1 eV, while thermal leptogenesis occurs naturally after a smooth exit from inflation into the radiation era. 
}

\begin{document}
\maketitle

\section{Introduction}

The Standard Model (SM) is an extremely successful quantum field theory framework that accurately describes all processes involving known elementary particles. Nevertheless, several crucial open problems remain unsolved within this framework and have motivated a plethora of extensions in the literature. 

On the one hand, several open problems are related to fine-tuning issues, such as the stability of the electroweak scale (the `hierarchy problem'), the smallness of the cosmological constant or the severe constraints on CP-violation in strong interactions. These issues can, however, only be seen at most as hints for new physics\footnote{Finely-tuned parameters are not a problem per se within a renormalizable quantum field theory, but only a symptom of their sensitivity to unknown physical properties at high energy scales.}. On the other hand, there are four problems that definitely point towards the existence of new particles and/or interactions: (1) neutrino masses, (2) dark matter, (3) the cosmological baryon asymmetry and (4) inflation. Neutrino masses cannot be accounted for with only the field content of the SM (see e.g.~\cite{Tanabashi:2018oca} for a review); there is mounting evidence for a particle nature of dark matter, as opposed to a modification of gravity (also reviewed in \cite{Tanabashi:2018oca}); CP-violation within the SM is manifestly insufficient to generate the observed matter-antimatter asymmetry (see e.g.~\cite{Cline:2006ts}); and standard cosmology fails to explain the nearly scale-invariant spectrum of Cosmic Microwave Background (CMB) anisotropies, inflation being the leading mechanism to generate this, alongside explaining the apparent fine-tuning of initial conditions embodied in the so-called horizon and flatness problems (see e.g.~\cite{Baumann:2009ds}).

While several `top-down' frameworks can address these problems, most predict a plethora of new physical states that have yet to be found in colliders or require too many unknown parameters to be fully tested experimentally. Arguably, a `bottom-up' approach of building minimalistic models, with as few novel ingredients as possible, may prove more fruitful from both the theoretical and experimental perspectives, keeping in mind that such models may find different embeddings in more complete theories, such as grand unification or superstring/M-theory. In fact, if suitable minimal models can be found to agree with experiment, this may provide further guidance into developing UV-complete theories. 

Hence, in this work, we focus on building a minimal quantum field theory model where the four problems singled out above can be simultaneously addressed. We elect the framework of {\it warm inflation} \cite{Berera:1995wh, Berera:1995ie} as our primary arena, since in this context interactions between the inflaton scalar field and other physical states play a relevant role in both the dynamics and observational consequences of inflation. This thus provides a natural paradigm within which to connect the high-energy physics relevant to describe the inflationary dynamics to low-energy observables and related open problems. 

The fundamental difference between warm inflation and the more conventional cold inflation models is the inclusion of dissipative effects that result from energy transfer between the inflaton field and an ambient radiation bath. These are non-equilibrium processes associated with particle production that, if significant, may counteract the diluting effect of accelerated expansion and keep the thermal heat bath at a slowly-varying temperature above the Hubble scale (which sets the Hawking temperature of the de Sitter horizon) (see \cite{Berera:2008ar} for a review). These effects also source thermal fluctuations of the inflaton field, which in turn produce a nearly scale-invariant spectrum of primordial curvature perturbations \cite{Taylor:2000ze, Hall:2003zp, Graham:2009bf, Ramos:2013nsa}. The shape of this spectrum, which is imprinted in CMB maps, depends on the form of the interactions between the inflaton field and particles in the thermal bath, which is the most interesting aspect of this alternative paradigm from a particle physics perspective.

A finite temperature during inflation is, however, challenging to realize in practice, due to potentially large thermal corrections to the inflaton's mass that could hinder the slow-roll dynamics, despite the additional dissipative friction \cite{BGR, YL}. The typically large classical values of the inflaton field also tend to make the particles it couples to quite heavy, greatly suppressing dissipative effects unless a large multiplicity of fields is present, which is a challenge in realistic extensions of the SM (see e.g.~\cite{BasteroGil:2009ec, BasteroGil:2011mr}). These problems have been recently overcome in the `Warm Little Inflaton' (WLI) scenario \cite{Bastero-Gil:2016qru}, where, as we detail below, collective symmetry breaking of a U(1) gauge symmetry and an additional discrete symmetry protect the flatness of the scalar potential against large thermal corrections and keep the mass of particles coupled to the inflaton field parametrically below the temperature of the heat bath (see also \cite{Bastero-Gil:2017wwl, Bastero-Gil:2018uep}). 

The WLI scenario also offers new ways to connect inflation to the other SM shortcomings outlined above. It has already been shown that the discrete interchange symmetry also makes the inflaton stable at late times. Since it only interacts directly with heavy particles, these interactions are only relevant at the high-temperatures attained during inflation. This implies that a cold, weakly interacting relic of the inflaton field remains until the present day and can account for the dark matter abundance \cite{Rosa:2018iff}\footnote{Such an inflaton remnant may also act as quintessence at late times \cite{Rosa:2019jci}, although this requires particular non-renormalizable forms of the potential that can only be justified within a more complete theory (see also \cite{Dimopoulos:2019gpz, Lima:2019yyv} for related scenarios).}.

The simple structure of the WLI Lagrangian is also quite suggestive of the particular implementation we seek in this work. The inflaton interacts with heavy fermions, which in turn may decay into a light scalar and a light fermion through standard Yukawa terms. It is thus natural to try to identify these heavy fermions with the right-handed neutrinos missing in the SM, and which commonly appear in unification theories based e.g.~on the gauge group SO(10). Their light decay products then correspond to the SM Higgs and lepton doublets. We will show that consistently realizing warm inflation within this setup requires Majorana masses for the right-handed neutrinos just below the GUT scale, at around $10^{15}$ GeV, and neutrino Yukawa couplings $\lesssim 1$. The standard type-I seesaw mechanism then naturally yields light neutrino masses $\lesssim$ 0.1 eV, in agreement with neutrino oscillation experiments. We will then explore in detail how the basic WLI structure can be modified to accommodate realistic neutrino masses and mixings by considering an additional spontaneously broken $\mathbb{Z}_3$ family symmetry.

Finally, the inclusion of Majorana right-handed neutrino singlets leads to the well-known thermal leptogenesis scenario in standard Big Bang cosmology \cite{Fukugita:1986hr}. We will show that our setup is no exception, and that this may occur, in particular, via the CP-violating decays of the lightest right-handed neutrino, which fall out-of-equilibrium after radiation has smoothly taken over the inflaton as the dominant component in the Universe. As we will see, the lightest right-handed neutrino species is not directly involved in the dissipative processes that keep the Universe warm during inflation, but obtaining a realistic neutrino mass spectrum at low-energies necessarily implies that these neutrinos are thermally produced before decaying after inflation.

This scenario constitutes the first successful implementation of warm inflation within a concrete extension of the SM, building upon a large body of work over the past decades in terms of understanding the intricacies of non-equilibrium processes during inflation \cite{Berera:1996nv, Berera:1996fm, Berera:1998px, Berera:1999ws, Berera:2001gs, Berera:2002sp, Berera:2004vm, Berera:2003kg, Moss:2006gt, BasteroGil:2006vr, Berera:2007qm, Moss:2008yb, Graham:2008vu, BasteroGil:2010pb, BasteroGil:2012cm, BasteroGil:2012zr, Bastero-Gil:2015nja, Berera:2018tfc, Das:2020lut}, model building \cite{Berera:1999wt, BasteroGil:2004tg, BasteroGil:2005gy, BuenoSanchez:2008nc, BasteroGil:2009gh, Sanchez:2010vj, Cai:2010wt, Cerezo:2012ub, Matsuda:2012kc, Bastero-Gil:2013owa, Li:2018riw, Bastero-Gil:2018yen}  and observational predictions \cite{Hall:2004ab, Moss:2007cv, Berera:2008ab, BasteroGil:2011xd, Bastero-Gil:2013nja, Bartrum:2013oka, Bartrum:2013fia, Bastero-Gil:2014jsa, Bastero-Gil:2014raa, Benetti:2016jhf, Benetti:2019kgw, Kamali:2019xnt}. Our construction further attests to the ability of the warm inflation framework to address several different open problems in both particle physics and cosmology \cite{Berera:1998hv, BasteroGil:2011cx, Bastero-Gil:2014oga, Bastero-Gil:2017yzb} and, as we will discuss, provides several experimental/observational imprints of its different aspects that would not be possible if the inflationary Universe were cold and empty as conventionally assumed.

This work is organized as follows. In the next section we review the warm inflation paradigm, discussing its dynamics and observational predictions, describing in detail the WLI scenario on which our model is based. We then introduce our basic particle physics model in section~\ref{sec:BasicIngredients}, highlighting the role of the different fields and symmetries involved. We analyze the cosmological dynamics associated with our scenario in section~\ref{sec:InflationaryDynamics}, first during inflation and then the post-inflationary evolution, in particular discussing how the inflaton may account for dark matter. Section~\ref{sec:NuMasses} is devoted to the spectrum of light neutrino masses, obtained through the seesaw mechanism, and here we explain how introducing a family-symmetry breaking flavon field can be accommodated within our basic setup. We use these results to show, in section~\ref{sec:Leptogen}, that thermal leptogenesis naturally occurs after inflation in this scenario. We summarize our main results in section~\ref{sec:Summary}, also discussing possible avenues for future research.


\section{Elements of warm inflation} \label{sec:WI}

Warm inflation (WI) is an inflationary paradigm where dissipative processes transfer energy from the inflaton scalar field, $\phi$, to light degrees of freedom in a nearly-thermal radiation bath. The cosmological dynamics of the inflaton-radiation system is described by the dynamical equations for the inflaton and radiation, alongside the Friedmann equation determining the Hubble expansion rate $H$:
\begin{gather}\label{EoMTime-init}
\ddot\phi + 3 H \dot\phi + \Upsilon \dot\phi + \partial_\phi V(\phi) = 0, \\
\dot\rho_r+4H\rho_r = \Upsilon \dot\phi^2, \\
H^2 = \frac{\rho_\phi + \rho_r}{3M_p^2}, \label{EoMTime-end}
\end{gather}
where $\rho_\phi =\dot\phi^2/2 + V(\phi)$ is the inflaton's energy density, $\rho_r \equiv (\pi^2/30) g_* T^4$ is the radiation energy density, with $g_*$ denoting the number of relativistic degrees of freedom, and $M_p \equiv \sqrt{\hbar c /(8 \pi G)} \approx 2.44 \times 10^{18} \text{ GeV}$ is the reduced Planck mass. The dissipation coefficient $\Upsilon=\Upsilon(\phi, T)$ can in general be computed using standard non-equilibrium quantum field theory techniques from the fundamental Lagrangian of a given particle physics scenario, at least close to thermal equilibrium with\footnote{This condition ensures that curved-space corrections may safely be neglected.} $T\gtrsim H$.

In the slow-roll approximation, the inflaton field and radiation fluid are slowly-varying: $\ddot \phi \approx 0$, $\dot \rho_r \approx 0$ while $\rho_\phi \approx V(\phi) \gg \rho_r$, therefore
\begin{gather}
\dot\phi \approx - \partial_\phi V(\phi)/(3 H(1+Q)), \label{SRphiEq}\\
\rho_r \approx \frac{3}{4}Q \dot\phi^2, \label{SRrhoEq}\\
H^2 \approx \frac{V(\phi)}{3M_p^2} \label{SRHEq},
\end{gather}
where $Q=\Upsilon/(3H) > 0$. This slow evolution occurs while the conditions:
\begin{equation}
\epsilon_\phi,~|\eta_\phi| < 1+Q,
\end{equation}
are valid, which generalizes the standard slow-roll conditions on the inflaton potential parameters $\epsilon_{\phi} \equiv (M_p^2/2) (\partial_{\phi}V/V)^2$ and $\eta_{\phi} \equiv M_p^2 (\partial_{\phi\phi}V/V)$. This means that the scalar potential needs not be as flat as in conventional models due to the additional dissipative friction.

Since inflation is a period of exponential expansion of the universe, it is more convenient to work with the number of e-folds of expansion, $N_e$, rather than the cosmological time. As such, Eq.~\eqref{SRphiEq} can be written in the form:
\begin{equation}\label{phi'/phi}
\frac{\phi'}{\phi} = -\frac{\sigma_\phi}{1+Q},
\end{equation}
where primes denote derivatives with respect to the number of e-folds and $\sigma_\phi =\! M_p^2 (\partial_{\phi}V/V\phi)$. 

Then, using Eqs.~\eqref{SRphiEq}-\eqref{SRHEq}, the ratio of radiation to inflaton potential energy density is given by;
\begin{equation}\label{RadDom}
\frac{\rho_r}{V}\approx \frac{1}{2}\frac{\epsilon_\phi}{1+Q}\frac{Q}{1+Q}.
\end{equation}
The radiation energy density thus remains a subdominant component ($\rho_r/V < 1$) while the slow-roll conditions hold, but may become comparable to the inflaton's energy density towards the end of inflation ($\epsilon_\phi \sim 1+Q$) if the system enters the strong dissipation regime\footnote{Note that $Q$ is a dynamical quantity, so that inflation may e.g.~start in the weak dissipation regime and end in the strong dissipation regime, or vice-versa.}, $Q> 1$. 

While the background homogeneous inflaton field evolves according to \eqref{EoMTime-init}, the full inhomogeneous field satisfies a Langevin-like equation with an additional Gaussian white noise term on the right-hand side. This essentially reflects the fluctuation-dissipation theorem, with the variance of the noise term being determined by the dissipation coefficient $\Upsilon$, and can be rigorously derived using non-equilibrium quantum field theory techniques. Although we refer the interested reader to the existing literature for further details (see e.g.~\cite{Berera:2008ar}), we point out that this is a fundamental property of any dissipative system. For instance, consider the Brownian motion of a particle in a gas - the average effect of random collisions with the gas molecules is a damping of the particle's motion (analogous to the effect of the $\Upsilon\dot\phi$ term in Eq.~\eqref{EoMTime-init}), but the particle's velocity can never truly reach zero due to the effect of individual collisions (which yields the random noise term).

This signals a fundamental difference between the cold and warm inflation paradigms. Whereas in the former inflaton fluctuations start in a quantum regime and then evolve towards a classical limit due to gravitational particle production in de Sitter space, in warm inflation the interactions with the heat bath source thermal inflaton fluctuations, which are born classical. The full power spectrum of primordial curvature perturbations generated by these thermal inflaton fluctuations has been computed in detail by several different works in the literature (see e.g. \cite{Ramos:2013nsa} and references therein), taking into account the dynamics of the fluctuations themselves (with the extra friction resulting in an earlier freezeout), the occupation numbers of inflaton particles in the thermal bath and the interplay between inflaton and radiation perturbations:
\begin{equation}\label{power_spectrum}
\Delta_\mathcal{R}^2 = \frac{V(\phi_*)}{24\pi^2 M_p^4}\frac{(1+Q_*)^2}{\epsilon_{\phi_*}}F(Q_*),
\end{equation}
where starred quantities are evaluated at the instant when fluctuations at the CMB pivot scale became super-horizon during inflation, generically 50-60 e-folds before it ended. The function $F(Q_*)$ is given by:
\begin{equation}\label{spectrumF}
F(Q_*) = \left(1+2n_*+\frac{2\sqrt{3}\pi Q_*}{\sqrt{3+4\pi Q_*}}\frac{T_*}{H_*}\right)G(Q_*),
\end{equation} 
where $n_*$ is the phase-space distribution of the inflaton fluctuations and $G(Q_*)$ is the correction due to the coupled evolution of inflaton and radiation fluctuations due to the temperature dependence of the dissipation coefficient. In general, this has to be computed numerically and depends on the form of $\Upsilon(T,\phi)$, as well as (mildly) on the scalar potential $V(\phi)$. For the case $\Upsilon\propto T$ (at least near-horizon crossing) and $V(\phi)\propto \phi^4$ that we will be interested in, a numerical fit gives $G(Q_*) \approx 1 + 0.0185 Q_*^{2.315} + 0.335 Q_*^{1.364}$ \cite{Bastero-Gil:2016qru}. 

The amplitude of the power spectrum $\Delta_\mathcal{R}^2 \approx 2.1\times 10^{-9}$ \cite{Akrami:2018odb} leads to a constraint on the magnitude of the inflaton potential, whereas the spectral index constrains the slow-roll parameters:
%
\begin{equation}
n_s-1 =\frac{d \log \Delta_\mathcal{R}^2}{d \log k} \simeq \frac{d \log \Delta_\mathcal{R}^2}{d N_e} \approx \frac{2 \eta_{\phi_*}-6\epsilon_{\phi_*}}{1+Q_*} \left(1-\frac{2 Q_*}{3+5Q_*} - \frac{Q_*(1+Q_*)}{3+5Q_*}\frac{\partial_{Q_*}F(Q_*)}{F(Q_*)} \right). \end{equation}
Note that this is similar to the spectral index in cold inflation, $n_s - 1 = (2 \eta_{\phi_*} - 6 \epsilon_{\phi_*})$, but multiplied by a $Q_*$-dependent factor which is smaller than $1$ (note that $\phi_*$ and $Q_*$ are related via Eq.~\eqref{RadDom}). Therefore, the curvature power spectrum is typically more blue-tilted in warm inflation than in cold inflation, although differences in the background evolution have to be taken into account.

Neither dissipative processes nor the finite temperature during inflation are, however, expected to source additional gravitational waves, given that the temperature is generically 3-4 orders of magnitude below the Planck scale. Hence, the enhancement of scalar curvature perturbations due to thermal effects typically leads to a suppression of the tensor-to-scalar ratio relative to cold inflation:
\begin{equation}\label{r-t}
r \equiv \frac{\Delta_{\text{t}}^2}{\Delta_\mathcal{R}^2} = \frac{16 \epsilon_{\phi_*}}{(1+Q_*)^2F(Q_*)}.
\end{equation}
We note again that, due to the difference in background evolution, the value of $\phi_*$ in cold and warm inflation is generically different for the same form of the inflaton potential. However, most warm scenarios are characterized by a low tensor-to-scalar ratio, even in the weak dissipation regime at horizon-crossing, $Q_*\ll 1$. This is an attractive feature of warm inflation, which allows, in particular, for an agreement between inflationary models with simple monomial potentials with the primordial spectrum inferred from CMB observations \cite{Bartrum:2013fia, Bastero-Gil:2016qru, Bastero-Gil:2017wwl, Bastero-Gil:2018uep, Bastero-Gil:2019gao}. This is not the case of the cold inflation scenario, where the simplest renormalizable monomials, $\phi^2$ and $\phi^4$, have been ruled out by the results of the Planck mission \cite{Akrami:2018odb}. 

In summary, warm inflation has several attractive features: an unobservable late reheating period is replaced by a smooth transition into a radiation-dominated epoch, with interactions between the inflaton field and other particles in the heat bath leaving an imprint on the primordial curvature spectrum; the additional friction facilitates slow-roll and the suppression of the tensor-to-scalar ratio brings the simplest models into agreement with observational data.


\subsection{The Warm Little Inflaton scenario}

Building successful realizations of warm inflation within quantum field theory remained, despite its appealing features, a challenge for more than two decades, since in general (i) the heat bath backreacts on the scalar potential, yielding in particular potentially large thermal mass corrections to the inflaton field that can spoil the slow-roll dynamics, despite the dissipative friction; (ii) particles in the thermal bath gain mass from the inflaton field, and may become too heavy to yield any significant dissipative effects. For instance, a Yukawa coupling of the form $g\phi\bar{\psi}\psi$ yields a mass $g\phi\gg T$ for the fermions in the thermal bath, as well as a thermal correction $\sim gT\gg H$ to the inflaton's mass, unless the Yukawa coupling $g\ll 1$, which would in turn also suppress the dissipation coefficient.

These problems were overcome, for the first time, in the `Warm Little Inflaton' (WLI) model\footnote{More recently, a realization of warm inflation based on an axion-like inflaton field with only derivative interactions with the thermal bath has been proposed in \cite{Berghaus:2019whh}.}, where in essence the inflaton field is coupled to two particle species in the thermal heat bath. The masses of these particles are oscillatory functions of the inflaton field, and hence bounded, and their combined effect leads to a cancellation of the leading thermal corrections to the scalar potential \cite{Bastero-Gil:2016qru}. 

The original model, on which we will base our subsequent construction, extends the SM by a U(1) gauge symmetry, under which two complex scalar fields $\Phi_1$ and $\Phi_2$ are equally charged. Both scalar fields acquire equal vacuum expectation values (VEVs) below a critical temperature and collectively break the U(1) symmetry. The vacuum manifold of the theory can then be parametrized as:
%
%
\begin{equation}\label{1phase}
\Phi_1 = \frac{1}{\sqrt{2}} (M + \rho_1) e^{i (\pi + \varphi)/(\sqrt{2}M)}, \qquad \Phi_2 = \frac{1}{\sqrt{2}} (M + \rho_2) e^{i (\pi -\varphi)/(\sqrt{2}M)},
\end{equation}
where $M$ sets the scale of spontaneous symmetry breaking. The overall phase $\pi$ can be removed, constituting the Goldstone boson that becomes the longitudinal component of the massive gauge boson. The remaining angular degree of freedom, $\phi$, then remains as a physical field that will act as the inflaton. Note that this corresponds to the relative phase between the two complex scalars, which does not change under U(1) gauge transformations. The inflaton is therefore a gauge singlet and the U(1) symmetry does not constrain its potential.

The complex scalars are also coupled, via standard Yukawa terms, to two additional fermions, $\psi_1$ and $\psi_2$, in the thermal bath. While the left-handed components of the fermions have the same charge as the complex scalars, their right-handed counterparts are gauge singlets. To avoid the `$\eta$-problem', i.e.~inducing large thermal corrections to the inflaton's mass, one imposes a discrete interchange symmetry, $\Phi_1 \leftrightarrow i \Phi_2$ and $\psi_1 \leftrightarrow \psi_2$, yielding an interaction Lagrangian of the form\footnote{For simplicity one works with $\phi \equiv \varphi/\sqrt{2}$, even though this is not the normalized field.}:
%
\begin{equation}
-\mathcal{L}_{\phi, \psi}=\frac{1}{\sqrt{2}} \left(g_1\Phi_1 + g_2\Phi_2 \right) \overline{\psi}_{1L}\psi_{1R} + \frac{1}{\sqrt{2}} \left(-ig_2\Phi_1 +ig_1\Phi_2 \right) \overline{\psi}_{2L}\psi_{2R} + h.c.. 
\end{equation}
If one chooses $g_1 = g_2 = g$, as in the original proposal in \cite{Bastero-Gil:2016qru}, the fermion masses after symmetry breaking are given by $M_1 = gM\cos{(\phi/M)}$, $M_2 = gM \sin{(\phi/M)}$. Since these are bounded functions, this allows the fermion fields to remain light throughout inflation and ensures non-negligible dissipation with only two fields. Alternatively, one can choose $g_{1,2} = g$ and $g_{2,1} = 0$ \cite{Rosa:2018iff}, in which case the fermion masses are independent of the inflaton field, $M_1 = M_2= gM/2$. In both cases, the combination $M_1^2 + M_2^2$ that determines the leading finite temperature correction to the scalar potential  is independent of the inflaton field, $\phi$, thus preventing the dangerous $\mathcal{O}(T^2)$ thermal mass corrections that can preclude slow-roll. In this work, we will considert he general case where $g_1 \neq g_2 \neq 0$, which exhibits in fact these very same features, as we show in detail in Appendix \ref{App:InflatonSelf-Energy}.

Note that the interchange symmetry leads to a reflection symmetry for $\phi$:
\begin{equation}\label{IntSym}
	\Phi_1 \leftrightarrow i \Phi_2 \quad \Longleftrightarrow\quad \phi \leftrightarrow \frac{\pi}{2}M - \phi,
\end{equation}
which in practice is a $\mathbb{Z}_2$ symmetry acting on $\phi/M-\pi/4$. Therefore, only interactions involving even powers of the (shifted) inflaton field are possible. This prevents the inflaton from decaying when close to the minimum at the origin, thus necessarily yielding a stable remnant at late times that can act as dark matter \cite{Rosa:2018iff}, as we will explore in more detail below in the context of the general model. Note also that the interchange symmetry requires equal charges and VEVs for $\Phi_1$ and $\Phi_2$, as assumed above.

With this basic setup in mind, we will now build a concrete extension of the SM where the fermions $\psi_1$ and $\psi_2$ are identified with two of the right-handed neutrinos.


\section{Basic particle physics setup}\label{sec:BasicIngredients}

Our concrete implementation of the WLI scenario is based, as mentioned above, on an identification of the $\psi_{1,2}$ fermions with two of the missing right-handed neutrinos in the SM, and which are singlets under the latter's gauge group. Naturally, the additional (gauged) U(1) symmetry can then be identified with lepton number or $B-L$, and generically we denote it by U(1)$_X$. Given the three fermion families in the Standard Model, it is natural to consider three additional right-handed neutrino Weyl fermions, $N_{1,2,3}$, as well as three additional complex scalar fields $\Phi_{1,2,3}$. However, the discrete interchange symmetry only acts on the fields in the first two generations, i.e.~$\Phi_1 \leftrightarrow i \Phi_2, N_1 \leftrightarrow N_2$. While it may a priori seem that the additional third family could be eliminated, we will see that the third right-handed neutrino state is crucial for obtaining realistic neutrino masses and mixings, with the VEV of $\Phi_3$ setting its Majorana mass. Including three right-handed neutrinos may also allow for a possible embedding into e.g.~an $SO(10)$ GUT.

Given the additional particle content required for a realistic model, compared to the original WLI setup, we also impose a discrete $\mathbb{Z}_3$ family symmetry that ensures that both $N_3$ and $\Phi_3$ do not affect the inflationary dynamics, in particular the cancellation of the leading thermal corrections to the inflaton's mass. In Table \ref{Tab:content}, we describe the particle content of the model and corresponding charge assignments for the gauge and flavour symmetry, with two possibilities in the latter case. Note that quarks are not directly involved in the inflationary dynamics, such that their charge assignments are left free. 
\begin{table}[h]
	\begin{center}
		\begin{tabular}{|c ||cc | cc | c | cccc |}
			\hline & & & & & & & & &  \\ [-2.5ex]
			& $\Phi_{1,2}$ & $\Phi_3$ & $N_{1,2}$ & $N_3$ & $H$ & ${L_{1,2}}$ & ${L_3}$ & ${\ell_{1,2}}$ & ${\ell_3}$  \\ [0.5ex] 
			\hline\hline
			$SU(3)_c$   & $1$ & $1$ & $1$ & $1$ & $1$ & $1$ & $1$ & $1$ & $1$  \\
			$SU(2)_L$   & $1$ & $1$ & $1$ & $1$ & $2$ & $2$ & $2$ & $1$ & $1$  \\
			$U(1)_Y$   & $0$ & $0$ & $0$ & $0$ & $\frac{1}{2}$ & -$\frac{1}{2}$ & -$\frac{1}{2}$ & -$1$ & -$1$ \\
			$U(1)_{X}$   & $2$ & $2$ & -$1$ & -$1$ & $0$ & -$1$ & -$1$ & -$1$ & -$1$  \\
			\hline
			$\mathbb{Z}_3$ (option I)   & $\omega$ & $\omega^2$ & $\omega$ & $\omega^2$ & 1 & $\omega$ & $\omega^2$ & $\omega$ & $\omega^2$  \\
			$\mathbb{Z}_3$ (option II)   & $\omega$ & $\omega^2$ & $\omega$ & $\omega^2$ & 1 & $\omega$ & $\omega$ & $\omega$ & $\omega$  \\
			\hline
		\end{tabular}
	\end{center}
	\caption{Field content of the model, for $X=B-L$. We give two optional charge assignments for the discrete $\mathbb{Z}_3$ family symmetry ($\omega = e^{2i\pi/3}$), differing only for ${L_3}$ and $\ell_3$.\label{Tab:content}}
\end{table}
We assume that all three complex scalar fields collectively break the $U(1)_X$ symmetry upon acquiring VEVs, with the interchange symmetry imposing equal VEVs for $\Phi_1$ and $\Phi_2$. The vacuum manifold is then given by:
\begin{align}
&\Phi_1 = \frac{1}{\sqrt{2}} (M+\rho_1)e^{i\pi/M_T}e^{i\xi\theta/ M_T}e^{i\phi/M} , \nonumber\\
&\Phi_2 = \frac{1}{\sqrt{2}} (M+\rho_2)e^{i\pi/ M_T}e^{i\xi\theta/ M_T}e^{-i\phi/M}, \\
&\Phi_3 = \frac{1}{\sqrt{2}} (M'+\rho_3) e^{i\pi/ M_T}e^{-2i\xi^{-1}\theta/ M_T},\nonumber
\end{align}
where $M_T^2=2M^2+M'^2$, $\xi=M'/M$ and in general $M\neq M'$. In this parametrization, $\pi$ denotes the Goldstone boson that is `eaten' by the $X$-gauge boson after spontaneous symmetry breaking, and which can be explicitly removed by going to the unitary gauge\footnote{We note that in the collective breaking of a U(1) symmetry by $n$ equally charged complex scalar fields with VEVs $\sqrt{2}\langle \Phi_i\rangle =v_ie^{i\varphi_i/v_i}$, $i=1,\ldots, n$, the Goldstone boson is the linear combination $\pi=\sum_i v_i\varphi_i /v$, where $v^2=\sum_i  v_i^2$.}. This leaves two physical gauge-invariant fields, $\phi$ and $\theta$. We take $\phi$ to play the role of the inflaton field, corresponding to the relative phase between the fields $\Phi_1$ and $\Phi_2$, as in the original WLI model. The interchange symmetry imposes only that the scalar potential for $\phi$ must be an even function of $\phi/M-\pi/4$, being otherwise arbitrary. The scalar potential for $\theta$ is arbitrary, being unconstrained by the interchange symmetry, and we assume, to simplify our analysis, that $\theta$ has a sufficiently large mass that we may set $\theta=0$ both during and after inflation.

We also assume, for simplicity, that the gauge coupling and scalar self-couplings are $\mathcal{O}(1)$, such that both the gauge field and the radial fields $\rho_{1,2}$ gain masses $\sim M$ and hence decouple from the dynamics at temperatures below the symmetry breaking scale at which inflation occurs. We may also neglect the radial field $\rho_3$, since its coupling to the inflaton field can be taken as arbitrarily small in a technically natural way.

The relevant interactions for the inflationary dynamics are those between the complex scalars and the right-handed neutrinos, which yield Majorana mass terms, alongside the Yukawa couplings involving $N_i$ and the SM Higgs and lepton doublets. The interchange and family symmetries then yield the general beyond the SM Yukawa Lagrangian density:
\begin{equation}\label{YukLag}
	\mathcal{L}_{\text{Yuk}} = \sum_{i,j} \left(G_{ij} \Phi_i \overline{N}_j^c N_j -   Y_{ij} \overline{N}_i H^\dagger L_j + \mathrm{h.c.}\right),
\end{equation}
with $Y_{1i}=Y_{2i}$ and
\begin{equation}
G=\begin{pmatrix} g_1 & g_2 & 0 \\ -i g_2 & i g_1 & 0 \\ 0&0&g_3 \end{pmatrix}.
\end{equation}
This structure implies that the inflaton field only interacts with two of the right-handed neutrinos, $N_{1,2}$, and only these are involved in the dissipative dynamics leading to warm inflation as we describe in the next section. The Yukawa terms allow these right-handed neutrinos to decay into SM leptons and Higgs doublet states (the Higgs boson itself and the components that will be `eaten' by the weak gauge fields after the electroweak phase transition). Dissipation of the inflaton's energy then occurs via the two-stage process $\phi \rightarrow N_{1,2} \leftrightarrow L H$ in a near-equilibrium regime. These are therefore the light degrees of freedom that make up the thermal bath during inflation, and later we will describe how the remaining SM states are thermally excited close to the end of inflation.

These very same interactions are also responsible for generating light neutrino masses at low energies through the type I seesaw mechanism \cite{Minkowski:1977sc}, such that:
\begin{equation} \label{eq-seesaw}
m_\nu = -m_D^T M_R^{-1} m_D,
\end{equation}
where $M_{Rj} =G_{ij}\langle\Phi_i\rangle$ is the diagonal Majorana mass matrix for the right-handed neutrinos and $m_D=v Y$ is the Dirac mass matrix, with $v\simeq 174$ GeV denoting the Higgs VEV after electroweak symmetry breaking. For a single family, this would yield $m_\nu\sim y^2 v^2 /M_R$. Taking this to match the largest mass splitting from solar neutrino oscillations, $m_\nu\sim 0.05$ eV \cite{Tanabashi:2018oca}, we find $M_R\sim 10^{14}-10^{15}$ GeV for $y\lesssim 1$ Yukawa couplings. As we will see in the next sections, this is precisely the parametric range required to consistently realize warm inflation.

We note that, by imposing the $\mathbb{Z}_3$ family symmetry, the third right-handed neutrino $N_3$ is effectively decoupled from the inflaton field, which is required to enforce the WLI mechanism for the cancellation of large thermal corrections to the inflaton's mass, which as described earlier and detailed in Appendix \ref{App:InflatonSelf-Energy} involves only two fields interacting with the inflaton. While this fixes the $\mathbb{Z}_3$ charge of the right-handed neutrinos, there is still an ambiguity in the charge assignments for the SM leptons. The first option in Table \ref{Tab:content} assigns the same $\mathbb{Z}_3$ charge for $N_3$ and $L_3,\ell_3$, effectively discriminating all the third family of leptons, which would allow for e.g.~an embedding in an $SO(10)$ GUT. In the second option, only $N_3$ has a different $\mathbb{Z}_3$ charge from the other leptons, effectively decoupling it from the low-energy theory. This leads to a two RH seesaw scenario, which is simpler and more predictive. We note that both options yield consistent realizations of warm inflation, although distinct low-energy phenomenology as we will explore below.


\section{Cosmological dynamics}\label{sec:InflationaryDynamics}

\subsection{Inflation}

To describe the dynamics of the inflaton field and radiation bath, we need to specify (i) the scalar potential for $\phi$ and its thermal corrections due to the right-handed neutrinos $N_1$ and $N_2$ and (ii) the dissipation coefficient associated with the production of the latter in the thermal bath by the slowly varying inflaton field.

As mentioned above, the inflaton is a gauge singlet and its potential function is arbitrary up to the reflection symmetry imposed by the underlying interchange symmetry. We note that this is very different from e.g.~an axion field, which is a (pseudo-)Goldstone boson and therefore inherits a shift symmetry from the underlying U(1) Peccei-Quinn symmetry. In our model, the Goldstone mode corresponds to the $\pi$ phase and not the inflaton field, which is invariant under U(1) transformations. Within an effective field theory approach, which underlies our `bottom-up' philosophy, we consider the most general renormalizable tree-level potential for $\phi$ including a quadratic and a quartic term:
\begin{equation} \label{InflatonPot}
V(\phi) =  \frac{1}{2}m_{\phi}^2 \left(\phi - {\pi\over4}M\right)^2 + \lambda \left(\phi - {\pi\over4}M\right)^4.
\end{equation}
During inflation, we will see that the field will take large background values $\phi\gg M$ for which the quartic term dominates and is well approximated by $\lambda\phi^4$, while at late times the field behaves as non-relativistic matter while oscillating close to the minimum, where the quadratic mass term dominates.

This potential is corrected at finite temperature due to the $\phi$-dependence of the $N_1$ and $N_2$ masses, which are given by\footnote{We use $N_{1,2}$ for the flavor and mass states interchangeably and distinguish between the two when necessary.}:
%
\begin{equation}\label{Masses}
\begin{aligned}
M_1^2&=&{M^2\over 4}\left[g_+^2\cos^2(\phi/M)+g_-^2\sin^2(\phi/M)\right],\\
M_2^2&=&{M^2\over 4}\left[g_+^2\sin^2(\phi/M)+g_-^2\cos^2(\phi/M)\right], 
\end{aligned}
\end{equation}
where $g_\pm= g_1\pm g_2$. Inflation will occur mostly in the regime where $T\gtrsim M_{1,2}$, where the high-temperature expansion of the effective potential is a good approximation and given by (per fermionic degree of freedom) \cite{Dolan:1973qd, Cline:1996mga}:
%
\begin{equation}
\begin{aligned}
V_{\text{high-T}}(T) =  
&T^4 \left[-\frac{7 \pi^2}{720} +\frac{1}{48} \frac{m^2}{T^2} + \frac{1}{64 \pi^2} \frac{m^4}{T^4} \left(\ln\left(\frac{m^2}{T^2}\right) - c_f \right) \right.  \\
&- \left. \frac{1}{2}\frac{m^2}{T^2} \sum_{l=2}^{n}\left(\frac{-m^2}{4\pi^2 T^2}\right)^l \frac{(2l-3)!! \zeta(2l-1)}{(2l)!!(l+1)} \left(2^{2l-1}-1\right)\right], \\
\label{HighTPot}
\end{aligned}
\end{equation}
where $c_f\simeq 2.635$. Since $M_1^2+M_2^2=M^2(g_+^2+g_-^2)/4$ is independent of the inflaton field, the leading thermal corrections to the inflaton's mass cancel out when adding the contributions of $N_1$ and $N_2$, which is the essence of the WLI mechanism that is therefore preserved in our implementation. An alternative way of showing this is to compute the thermally corrected inflaton self-energy in the high temperature regime, a computation we  describe in detail in Appendix \ref{App:InflatonSelf-Energy}. We note that quadratic divergences in the quantum loop-corrections to the inflaton's mass also cancel for the same reason, which is reminiscent of the `Little Higgs' mechanism to address the electroweak hierarchy problem.

There remain sub-leading thermal corrections of the Coleman-Weinberg form, which add up with the corresponding zero-temperature corrections. Remarkably, the two contributions add up in such a way that the argument of the logarithm in Eq.~(\ref{HighTPot}) above becomes simply $\mu^2/T^2$, where $\mu$ is the renormalization scale. It is therefore natural to take this close to the temperature of the thermal bath, for instance at horizon-crossing. Since the temperature varies slowly during inflation, the magnitude of this contribution is suppressed and, more importantly, it only induces oscillatory contributions to the slow-roll parameters, which have a negligible effect as described in the original WLI proposal \cite{Bastero-Gil:2016qru}.

Since we will consider the post-inflationary evolution, when the temperature falls below the mass scale of the right-handed neutrinos, we must consider also the low-temperature approximation of the effective potential (per fermionic degree of freedom)\cite{Dolan:1973qd,Cline:1996mga}:
%
\begin{equation}
V_{\text{low-T}}(T) = - T^4 e^{-m/T} \left(\frac{m/T}{2 \pi}\right)^{3/2} 
 \sum_{l=0}^{t} \frac{1}{2^l l!}\frac{\Gamma(l+5/2)}{\Gamma(l - 5/2)} \left(\frac{T}{m}\right)^l, \label{LowTPot}
\end{equation}
where the Boltzmann-suppression of thermal effects is manifest. In our numerical simulations we use both forms of the effective potential matched in such a way (similar to Sec. 2.2. of \cite{Cline:1996mga}) that a precise description of the full effective potential is obtained. We find that using $n=3$ (high-$T$) and $t=3$ (low-$T$), with the latter turned on at $T/M_1 \sim 0.5$ provides a smooth enough transition, yielding an error around $1\%$ (see also e.g.~\cite{Borges:2016nne}).

The thermal dissipation coefficient $\Upsilon$ can be computed using non-equilibrium quantum field theory techniques at finite temperature, provided that (i) the temperature exceeds the Hawking temperature of de Sitter space $H/2\pi$, so that the computation can be done in flat space; (ii) the thermal bath remains close to thermal equilibrium, which requires equilibration processes to occur faster than expansion; and (iii) the inflaton's dynamics is `adiabatic' compared to the thermal processes, which is justified within the slow-roll approximation provided that the latter condition is satisfied. Physically, the dissipative friction will correspond to a non-equilibrium production of right-handed neutrinos $N_1$ and $N_2$, which are kept close to equilibrium via their decays into light leptons and Higgs particles (including the would-be $W$ and $Z$ longitudinal modes). The above conditions may then be satisfied in the regime $T\gtrsim \Gamma_{N_{1,2}}\gtrsim H$, which will constrain the parametric regimes where warm inflaton can successfully be realized.

We detail the computation of the dissipation coefficient in Appendices \ref{AppB:FDW} and \ref{AppC:DissCoef}, where in the former we describe the computation of the thermal mass and decay width of the  right-handed neutrinos, which are needed to compute $\Upsilon$. For clarity, here we give only the approximate expressions for the contribution of each right-handed neutrino to the dissipation coefficient in the high and low-temperature regimes\footnote{This is the leading result in the narrow width approximation for the right-handed neutrinos, which vanishes if only one of the couplings is non-zero (i.e.~$g_-=\pm g_+$). In the latter case one needs to go beyond this leading approximation, as shown in \cite{Rosa:2018iff}.}:
\begin{align}
\Upsilon_i =&\frac{ \left( g_1^2 +g_2^2 \right)\cos^2\delta}{4 y^2}  \frac{\pi}{9-6\ln{(M_i/T)}}T, \quad T\gtrsim M_i \\ 
\Upsilon_i =&\frac{ \left( g_1^2 +g_2^2 \right)\cos^2\delta}{2 y^2}\left(\!{2M_iT\over \pi}\right)^{\!1/2}\!\!e^{-{M_i\over T}} , \quad T\ll M_i \label{LowTDissCoef}
\end{align}
for $i=1,2$, where we have defined the effective neutrino Yukawa coupling $y^2=\sum_{j=1}^3 Y_{1j}^2=\sum_{j=1}^3 Y_{2j}^2$ and the inflaton-neutrino `mixing angle':
\begin{equation}
-\delta= \tan^{-1}\left({g_-\over g_+}\tan(\phi/M)\right)+\tan^{-1}\left({g_-\over g_+}\cot(\phi/M)\right)~.
\end{equation}
We note that in the expressions above it is implicit that the right-handed neutrino masses include thermal corrections, although for simplicity we used the same notation as for the zero-temperature inflaton-dependent masses. Furthermore, we use, in our numerical simulations, the more complete expressions given in Appendix \ref{AppC:DissCoef}, that are suitable to explore the regime $T\sim M_i$. 

In the high-temperature regime relevant for inflation, the dissipation coefficient is essentially proportional to the temperature, as in the original WLI model\footnote{There is an oscillatory dependence on the inflaton field through the mixing angle $\delta$, but for large field excursions $\Delta\phi\gg M$ we may take an averaged value of $\cos^2\delta$.}. In the opposite limit dissipative effects become Boltzmann-suppressed\footnote{There are still residual contributions from virtual right-handed neutrino modes, but these are too suppressed to have any significant impact on the cosmological dynamics.}, and the right-handed neutrinos effectively decay away. As we will see below and originally shown in \cite{Rosa:2018iff}, this leaves a stable relic inflaton field that interacts very weakly with SM particles, and is therefore a natural dark matter candidate.

We now have everything in place to study the dynamics of the inflaton-radiation system, and it is convenient to rewrite Eqs.~\eqref{EoMTime-init}-\eqref{EoMTime-end} in terms of the number of e-folds of expansion, $dt = H dN_e$, 
\begin{gather}
\phi'' + 3 \left(1+ \frac{H'}{H} + \frac{\Upsilon}{H} \right) \phi' +  \frac{\partial_\phi V(\phi)}{H^2} = 0, \\
\rho_r' + 4\rho_r = \Upsilon H \phi'^2, \\
H^2 = \frac{\rho_\phi + \rho_r}{3M_p^2}. 
\end{gather}
Note that the radiation energy density can be computed from the contribution of all light particles to the thermal effective potential given above via $\rho_r = V_T + T s$ with $s = - dV_T/dT$. In the high-temperature regime, this yields at leading order the well-known expression $\rho_r\simeq (\pi^2/30)g_* T^4$, but this allows us to accurately track the contribution of $N_{1,2}$ to $g_*(T)$ as the temperature falls below their mass threshold.

\begin{figure}[htb!]
	\centering\includegraphics[width=0.6\textwidth]{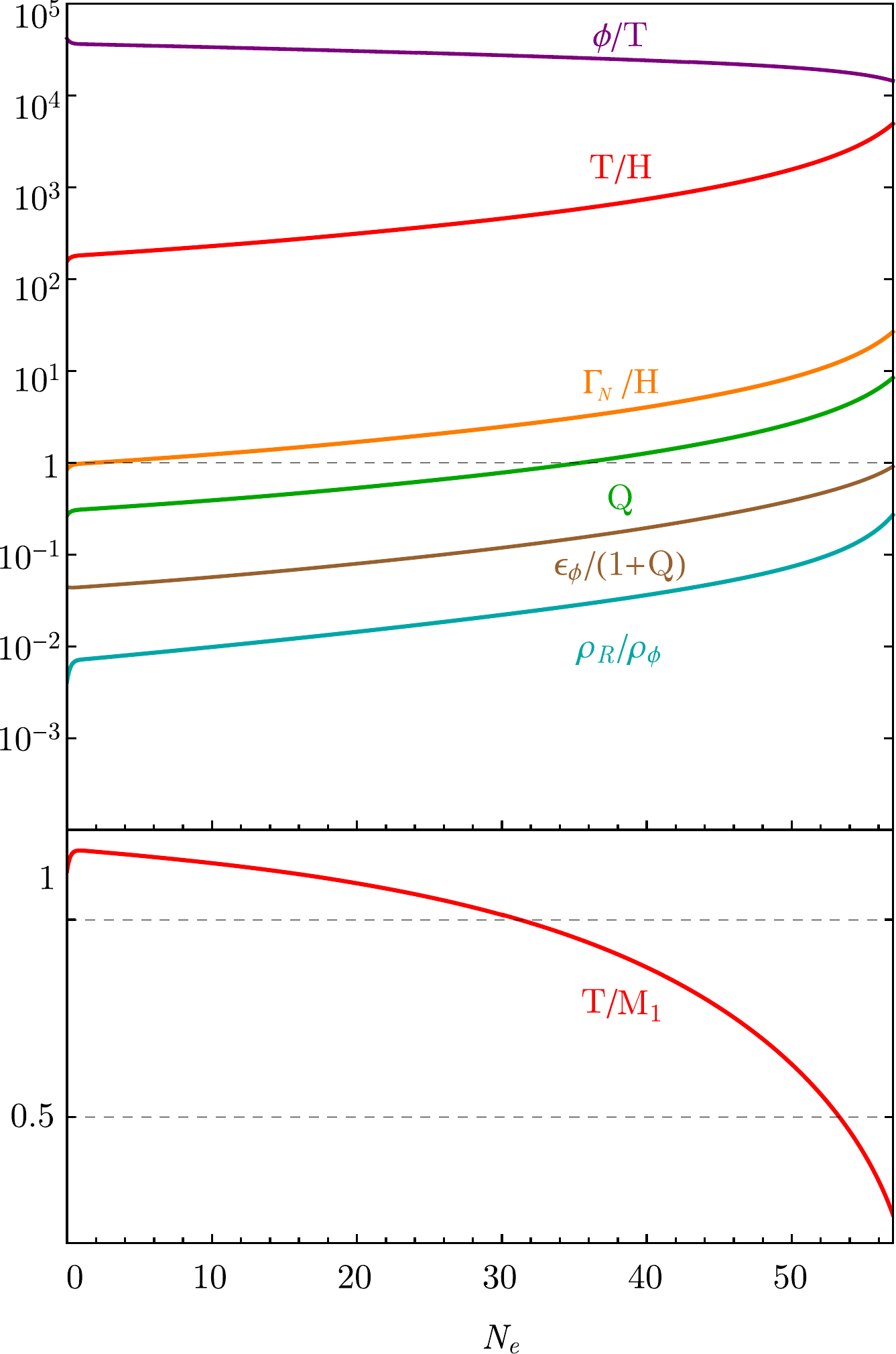} 
	\caption{Inflationary dynamics for $V(\phi)=\lambda \phi^4$ as a function of the number of e-folds. The total number of e-folds of inflation is $N_{\text{eq}} \approx 58$ with $(n_s,r)=(0.967,~2.6\times 10^{-4})$ (nearly-thermal inflaton fluctuations). The number of relativistic degrees of freedom during inflation is approximately constant, $g_* \approx 19$. The `reheating' temperature is $T_R \approx 7.5 \times 10^{13}$ GeV. Parameters: $Q_* = 0.8,~\phi_* \simeq12 M_{p},~ \lambda \approx 5 \times 10^{-16}$ ($y=0.74,~g_1=0.21, g_2 = 0.17,~M \approx 3.3 \times 10^{15}$ GeV).  \label{inflationary_dynamics} } 
\end{figure}

The above system of equations can be solved analytically in the high-temperature, slow-roll regime, where $\Upsilon\propto T$ and $V(\phi)\propto \phi^4$. However, since the temperature of the thermal bath is close to the masses of the right-handed neutrinos in the interesting parametric regimes, it is more accurate to numerically solve the above dynamical system using the full forms of the dissipation coefficient and effective potential with minimal approximations. In figure~\ref{inflationary_dynamics} we give a representative example of the dynamics of warm inflation in our model. 

As one can see, the system remains in a warm ($T> H$), near-equilibrium ($\Gamma_N> H$), slow-roll ($\epsilon_\phi<1+Q$) regime for nearly 60 e-folds of accelerated expansion. At horizon-crossing, dissipative effects are relatively weak $Q_*\lesssim 1$, but become sufficiently strong towards the end of inflation for radiation to become a sizeable component and soon come to dominate over the inflaton field, thus smoothly ending inflation (see also figure~\ref{cosmological_dynamics}). We find that the temperature at which radiation takes over, which constitutes an effective `reheating temperature' although the Universe never cooled down during inflation, is typically $T_R\sim 10^{13}-10^{14}$ GeV.

We note that accelerated expansion only truly ends when $\rho_r\approx \rho_\phi$, since $\ddot{a}\propto 2(\rho_\phi-\rho_r) $, which typically takes place 2-3 e-folds after the slow-roll conditions fail. Hence, the example in figure~\ref{inflationary_dynamics} has $\simeq 60$ e-folds of accelerated expansion.

Although the temperature (bottom red curve in figure~\ref{inflationary_dynamics}) remains close to the right-handed neutrino masses throughout inflation, it falls significantly below it at the end of the slow-roll phase, which in turn shuts down dissipative effects due to Boltzmann suppression as discussed above. Once dissipation can no longer counteract the diluting effect of expansion at this stage, the temperature of the thermal bath simply redshifts as $T\propto a^{-1}$. One must also take into account that other SM degrees of freedom also start being thermally excited towards the end of inflation as the ratio $T/H\sim 10^{4}$ and typical gauge processes start competing with the expansion rate. Consequently, the temperature falls by an additional factor $ (g_{*,\text{inf}}/g_{*, \text{SM}})^{1/3} \sim 2$, which also speeds up the shut down of dissipation at the end of inflation.

Inflaton particles are also produced in the thermal bath via several processes, including e.g.~right-handed neutrino annihilation and Landau damping processes, as well as lepton-Higgs annihilation via virtual right-handed neutrinos. The rate of each individual process is typically suppressed compared to the right-handed neutrino decay rate, since $g_{1,2}< y$ is required to have sufficiently light right-handed neutrinos. Nevertheless, the number of different processes producing inflaton particles in the thermal bath should partially compensate for this, such that inflaton particle production may compete with Hubble expansion already at horizon-crossing. This has implications for observational predictions, since inflaton occupation numbers at horizon crossing influence the shape of the primordial spectrum of perturbations, namely through the factor $2n_*$ in Eq.~\eqref{spectrumF}, with $n_*\sim T_*/H_*$ for a nearly-thermal distribution. Since a detailed analysis of inflaton particle production rates is beyond the scope of this work, we consider the two limiting regimes $n_*\ll 1$ and $n_*\simeq T_*/H_*$ in computing the spectrum of primordial perturbations. 

\begin{figure}[h!]
	\centering\includegraphics[width=0.6\textwidth]{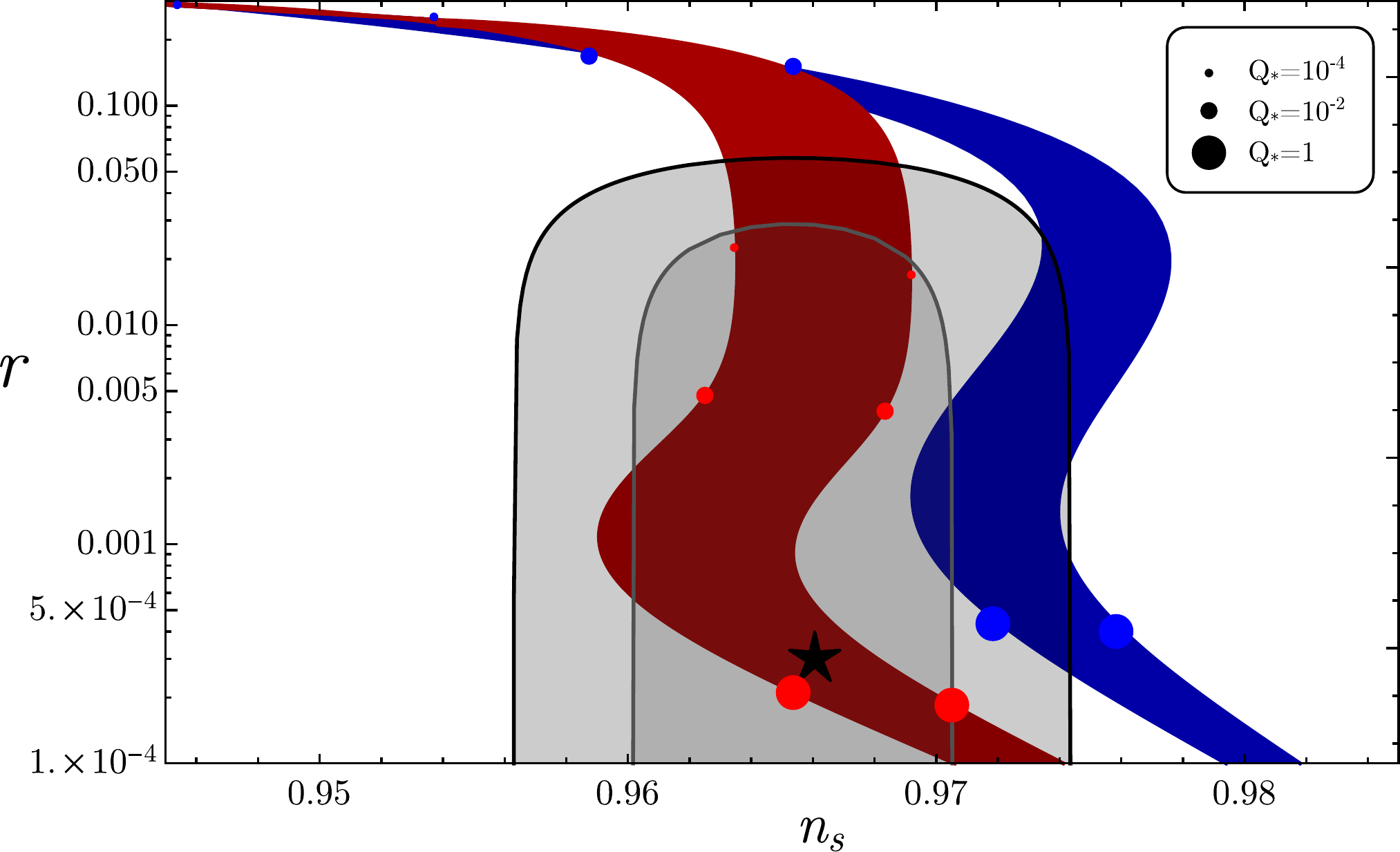} 
	\caption{Inflationary spectrum ($n_s,~r$) for $V(\phi)=\lambda \phi^4$ as a function of $Q_*$, with the $r$ axis in $\log$-scale. The red (blue) shaded region corresponds to the case where the inflaton fluctuations have a nearly-thermal (negligible)  spectrum at horizon crossing. The left (right) contour of these regions correspond to a $N_e = 50$ ($60$) of inflation. As $Q_* \rightarrow 0$, both regions converge to a straight line encompassing the predictions of cold inflation, $(n_s,~r) \in (0.94-0.95,0.32-0.27)$ for $N_e = 50-60$. The black star represents the case of figure \ref{inflationary_dynamics}. The black (gray) contour signals the $95\%$ ($68\%$) confidence limit of \cite{Akrami:2018odb} for the TT,TE,EE+lowE+lensing +BK14+BAO dataset.} \label{inflationary_spectrum} 
\end{figure}

In figure~\ref{inflationary_spectrum} we show the observational predictions for the curvature spectral index $n_s$ and tensor-to-scalar ratio $r$ in our model as a function of the dissipative ratio at horizon-crossing, $Q_*$ (with other dynamical variables fixed by the number of e-folds of inflation and the amplitude of the primordial curvature perturbation spectrum), in these two limiting regimes. 

These results are analogous to the original WLI scenario, which also considered a quartic potential for the inflaton field. 
As one can see, the spectrum is more blue-tilted than the corresponding cold inflation case, which is essentially a consequence of the thermal nature of inflaton fluctuations and of the associated growth of the dissipative ratio $Q$ and of the ratio $T/H$ during inflation, with modes leaving the horizon later during inflation having a (slightly) larger amplitude. Alongside the suppression of the tensor-to-scalar ratio typical of warm models, this makes the quartic inflaton potential agree with observational data from Planck for $Q_*\sim 10^{-2}-1$, which is quite remarkable since in the absence of dissipative effects this model has been completely ruled out, as first pointed out in \cite{Bastero-Gil:2016qru}. For $Q_*\gtrsim 1$ the interplay between inflaton and radiation fluctuations leads to the appearance of a growing mode in the spectrum, parametrised by $G(Q_*)$ in \eqref{spectrumF}, which renders the spectrum too blue-tilted. One should bear in mind that such a growing mode may potentially be damped by additional viscous effects in the thermal bath \cite{BasteroGil:2011xd}, such that $Q_*\gtrsim 1$ scenarios may also be in agreement with observations. Such a possibility is, however, still the object of an ongoing discussion so we will not pursue it any further, and restrict our analysis below to the regime where $Q_*\lesssim 1$.

Another distinctive observational signature of warm inflation that is also present in our scenario is non-Gaussianity, with a particular bispectral shape. For $\Upsilon\propto T$, $|f_{NL}^{\mathrm{warm}}|\lesssim 10$ for the range of $Q_*$ in agreement with observations mentioned above \cite{Moss:2007cv, Bastero-Gil:2014raa}. Dedicated searches made by the Planck collaboration for the `warm shape' of the bispectrum are, however, still far from reaching the required sensitivity, allowing for $Q_* < (3.2 - 4)\times10^3$ (95\% C.L.) \cite{Ade:2015ava, Akrami:2019izv}, which we hope may be improved with future experiments.

Finally, we note that one of the appealing features of warm inflation is the elimination of the uncertainty in the number of e-folds of inflation after horizon-crossing of the relevant CMB scales, due to the fast and smooth transition into a radiation-dominated universe, as recently observed in \cite{Das:2020xmh}. In fact, one may argue that warm and cold inflation involve a comparable number of free parameters, since interactions between the inflaton and other fields are necessarily present in both inflationary paradigms. In warm inflation these determine the full dynamics of accelerated expansion, whereas in cold inflation they only affect the reheating period and therefore result in an uncertainty in the total number of e-folds. In figure~\ref{inflationary_spectrum} we nevertheless included a $50-60$ e-folds uncertainty to accommodate possible non-standard cosmic histories after inflation. For instance, one or more short periods of thermal inflation \cite{Lyth:1995hj, Lyth:1995ka}\footnote{Note that thermal and warm inflation are distinct scenarios, since in the former there is no dissipative dynamics sustaining the temperature of the thermal bath and the inflaton is held by thermal effects at a metastable minimum. In fact, the latter correspond to the very same thermal mass corrections that need to be suppressed to realize warm inflation in the slow-roll regime.} may be necessary to dilute any unobserved relics potentially produced after warm inflation, given the large temperatures attained.


\subsection{Inflaton dark matter} \label{ssec:PostInf}


%
As we have seen above, once inflation ends dissipative effects quickly shut down and the inflaton field becomes underdamped and starts oscillating about the minimum of the potential. Since the potential is still dominated by the quartic term following inflation, the oscillating inflaton field first behaves as a dark radiation component, with an amplitude $\phi\propto a^{-1}$. This phase lasts until the amplitude of the inflaton field drops below $\phi_{\text{DM}} = m_{\phi}/\sqrt{2 \lambda}$, after which the quadratic term of the potential becomes dominant and the inflaton remnant behaves as non-relativistic (pressureless)  matter, with $\phi\propto a^{-3/2}$. 
\begin{figure}[h!]
	\centering\includegraphics[width=.6\textwidth]{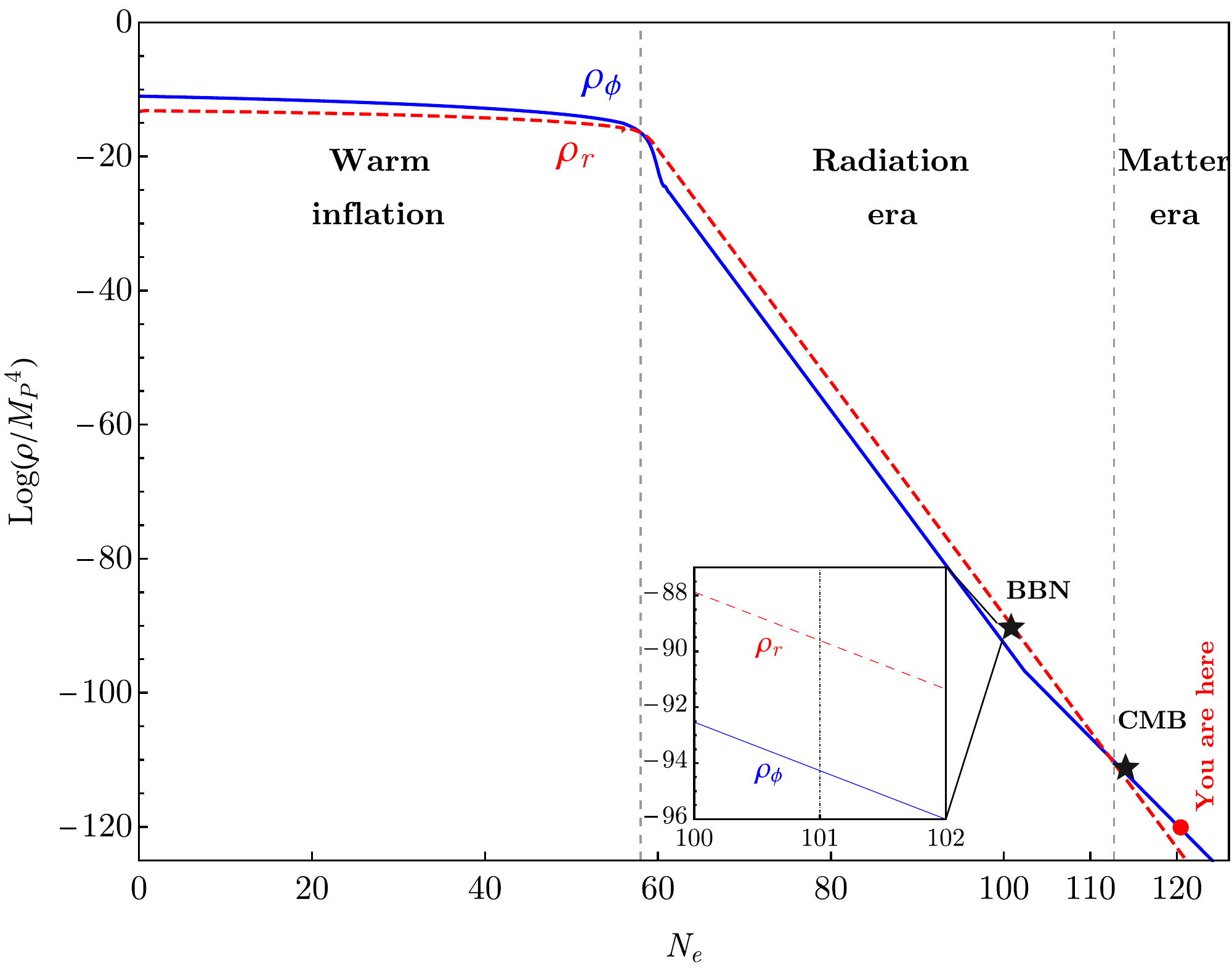} 
	\caption{Cosmological dynamics. The parameters are identical to those of figure \ref{inflationary_dynamics}. The dashed lines represent the two inflaton-radiation equalities. Big bang nucleosynthesis (BBN) ($T \sim 1$ MeV) and the last scattering surface (CMB) ($T \sim 0.3$ eV) are also highlighted. The inset shows a close-up of the energy densities during BBN, where $f \equiv \rho_{\phi}/\rho_r \sim 10^{-5}$.} \label{cosmological_dynamics}
\end{figure}
Due to the underlying interchange symmetry, which as discussed earlier corresponds to a $\mathbb{Z}_2$ reflection symmetry, the inflaton field becomes stable and, moreover, since the temperature falls below the right-handed neutrino mass threshold, the inflaton can only interact with SM particles via the exchange of virtual right-handed neutrinos. The inflaton field thus becomes a cold, weakly interacting relic after inflation, making it a natural dark matter candidate. 

Scattering processes mediated by virtual right-handed neutrinos may nevertheless lead to the thermalization of the oscillating inflaton condensate, in which case the inflaton would become a WIMP-like candidate that would eventually decouple from the thermal bath. The analysis made in \cite{Rosa:2018iff} for a generic WLI model shows, however, that the large right-handed neutrino masses $M_1=M_2=\sqrt{g_1^2+g_2^2}M/2$ greatly suppress such processes, which therefore do not significantly affect the field's dynamics\footnote{The results of \cite{Rosa:2018iff} can be immediately translated to our model with the appropriate identification of the generic scalar and fermion fields considered there with the right-handed neutrinos, leptons and Higgs fields.}. The oscillating inflaton field may nevertheless produce $\phi$ particles through the $\lambda\phi^4$ self-interaction while in the quartic regime. From the point of view of the cosmological dynamics, such inflaton particles are indistinguishable from an oscillating field since they do not thermalize and behave first as dark radiation and then as cold dark matter. As argued in \cite{Rosa:2018iff}, the transition between these two regimes occurs essentially at the same time as for the oscillating field, since they are produced with a momentum $p\sim \sqrt{\lambda} \phi$ in the dark radiation phase, which redshifts exactly like the amplitude of the oscillating field in the quartic potential. The amount of expansion required to reach the non-relativistic regime $p\lesssim m_\phi$ thus corresponds to the amount needed for $\phi\lesssim \phi_{\text{DM}}$.

After inflation, there are three distinct contributions to the energy density of the Universe, including (i) the dominant radiation composed by relativistic SM particles, (ii) the remnant background inflaton field that behaves initially as dark radiation while the field oscillates about the quartic potential, and (iii) inflaton particles in the thermal bath, which thermalized with the remainder relativistic degrees of freedom during inflation and decoupled from the latter once the right-handed neutrinos $N_{1,2}$ became non-relativistic. Note that thermalization of inflaton particles should occur before the end of inflation even if not already at horizon-crossing as discussed above, given that the $g_{1,2}$ couplings are not too suppressed in parametric regimes where the strong dissipation regime is reached before inflation ends. 

The fractional contribution of the decoupled inflaton particles is given by:
\begin{equation}
\frac{\rho_{\delta \phi}}{\rho_r}(T) = \frac{1}{g_{*, \text{SM}}} \left[\frac{g_*(T)}{g_{*, \text{SM}}}\right]^{1/3},
\end{equation}
where we have used entropy conservation, while the background inflaton remnant, in its dark radiation regime, contributes as: 
\begin{equation}
\frac{\rho_{\phi}}{\rho_r} (T) = f  \left[\frac{g_*(T)}{g_{*, \text{SM}}}\right]^{1/3},
\end{equation}
where $f$ is the ratio between the background inflaton and the radiation energy densities at the onset of field oscillations, which remains constant apart from temperature jumps due to varying $g_*$. Numerically, we find $f\sim 10^{-(4-5)}$ in the parametric range of interest for a successful realization of warm inflation. 

Both these inflaton components will contribute to the present dark matter abundance, and to assess which is the dominant one, we note, on the one hand, that the background remnant starts behaving as dark matter (i.e. the quadratic term in the potential becomes dominant) for field amplitudes $\lesssim \phi_{\text{DM}}= m_{\phi}/\sqrt{2\lambda}$. This corresponds to a temperature:
\begin{equation}
T_{\mathrm{DM}} = m_{\phi} \left( \frac{30}{2\pi^2 \lambda f} \right)^{1/4} {g_{*, \text{SM}}^{1/12}\over g_*(T_1)^{1/3}}~.
\end{equation}
On the other hand, the decoupled inflaton particles behave as dark radiation for longer (and hence are more diluted) than the background remnant, as they only become non-relativistic at $T=m_\phi \lesssim T_{\mathrm{DM}}$, yielding for $T<m_\phi$:
\begin{equation}
\frac{\rho_{\delta \phi}}{\rho_\phi} \simeq  6\times 10^{-4} \left(\frac{10^{-4}}{f}\right)^{3/4}\left(\frac{\lambda}{10^{-16}}\right)^{1/4} g_*(T_1)^{1/3},
\end{equation}
so that the background remnant is generically the dominant dark matter component. Then, using the present value of the dark matter abundance, we find for the inflaton mass and temperature at which the remnant starts behaving as cold dark matter:
\begin{equation}
\begin{aligned}
m_{\phi} &\approx \left(\frac{\Omega_c}{0.25}\right) \left(\frac{g_*}{106.75}\right)\left(\frac{10^{-15}}{\lambda}\right)^\frac{1}{2}\left(\frac{10^3}{\phi/T}\right)^{3} \text{ eV}, \\
T_{\text{DM}} &\approx 0.11 \left(\frac{\Omega_c}{0.25}\right) \left(\frac{g_*}{106.75}\right)^\frac{4}{3}\left(\frac{10^{-15}}{\lambda}\right)\left(\frac{10^3}{\phi/T}\right)^{4}  \text{MeV},
\end{aligned}
\end{equation}
where $\Omega_c$ is the relative dark matter density present today. For the example depicted in figures~\ref{inflationary_dynamics} and \ref{cosmological_dynamics}, $m_{\phi} \approx 0.4\  \text{eV}$ and $T_{\text{DM}} \approx 40 \text{ keV}$. We note that one should take into account the radiative corrections from the two right-handed neutrinos to the inflaton's mass to assess whether such small values of the inflaton mass require fine-tuning. In particular, one would expect Coleman-Weinberg contributions to the inflaton potential $\sim {M_i^4/64\pi^2}$ from both $N_1$ and $N_2$. However, it is well known that this standard form of the Coleman-Weinberg potential is obtained via mass-independent renormalization schemes, such as $\overline{MS}$, which do not respect the Appelquist-Carazonne decoupling theorem \cite{Appelquist:1974tg}. A computation of the 1-loop effective potential (at zero temperature) using a mass-dependent renormalization scheme gives a modified form of the Coleman-Weinberg term, generically given by \cite{BasteroGil:2010vq}:
\begin{equation}
\Delta V^{(1)}= \sum_i (-1)^F{m_i^4\over 64\pi^2}\left[\log\left({m_i^2\over \mu^2}\right)-I\left({m_i^2\over \mu^2}\right)\right]~,
\end{equation}
where $\mu$ is the renormalization scale, the sum is over all bosons and fermions with masses $m_i$, and
\begin{equation}
I(x)=\log(x)-2-\sqrt{1+4x}\log\left({\sqrt{1+4x}-1\over \sqrt{1+4x}+1}\right)~. 
\end{equation}
While for $\mu\gtrsim m_i$ this yields the standard Coleman-Weinberg expression, this is no longer the case at low energies $\mu\ll m_i$. Thus, for $\mu\ll M_{1,2}$ the leading contribution of the right-handed neutrinos to the effective potential is $\Delta V^{(1)}_{N_{1,2}}= \mu^2 (M_1^2+M_2^2)/192\pi^2+\mathcal{O}(\mu^4)$. As we have seen above, this is independent of the inflaton field, regardless of the specific choice of $\mu$ in this limit. Thus, the right-handed neutrinos do not generate large radiative corrections to the inflaton's mass at low energies/temperatures, through exactly the same mechanism that cancels large thermal corrections in the high-temperature limit. Hence, no fine-tuning of the inflaton mass is needed to obtain the observed dark matter abundance within the proposed setup.

As originally discussed in \cite{Rosa:2018iff}, the fact that the inflaton field plays a role in the post-inflationary Universe, and in particular naturally accounts for all dark matter, yields specific observational signatures in addition to the warm inflation observables discussed in the previous section. 

First, inflaton fluctuations on super-horizon scales generate not only curvature perturbations but also perturbations in the dark matter density itself, since $\rho_\phi\propto \phi^2$ \footnote{Here we have shifted the inflaton field value $\phi$ by $M\pi/4$ without loss of generality.}. Hence, the model predicts cold dark matter isocurvature modes that are correlated with the leading adiabatic curvature fluctuations. These can be parametrized via  \cite{Lyth:2002}:
\begin{equation}
S_c = -3 H \left(\frac{\delta \rho_c}{\dot{\rho}_c} - \frac{\delta \rho_r}{\dot{\rho}_r}\right)= \frac{\delta \rho_c}{\rho_c} - \frac{3}{4}\frac{\delta \rho_r}{\rho_r}= 2 \frac{\delta \phi}{\phi} - 3 \frac{\delta T}{T}~.
\end{equation}
Since, during warm inflation with $\Upsilon\propto T$ as in our scenario, we have at horizon-crossing:
\begin{equation}
\frac{\delta T_*}{T_*} = \frac{4Q_*}{3+5Q_*}\frac{\delta \phi_*}{\phi_*}~,
\end{equation}
we find for the ratio between isocurvature and adiabatic modes:
\begin{equation}\label{iso/adi ratio}
B_c = \frac{S_c}{\mathcal{R}} =  -\left( 2 - \frac{12 Q_*}{3+5Q_*} \right)\frac{\sigma_{\phi_*}}{1+Q_*}~,
\end{equation}
where we have used that $\mathcal{R} = H (\delta \phi/\dot\phi)=\delta\phi/\phi'$ and Eq.~\eqref{phi'/phi}. Since the primordial spectrum of curvature perturbations in our setup requires $Q_* < 3$, we conclude that cold dark matter isocurvature perturbations are anti-correlated with the main adiabatic component. 

The Planck collaboration has placed stringent bounds on the magnitude of such modes, typically parametrized in terms of $\beta_{\text{Iso}}=B_c^2/(B_c^2 + 1)$. Using the values of $Q$ and $\phi$ at horizon crossing of the example in figures~\ref{inflationary_dynamics} and \ref{cosmological_dynamics}, we obtain $\beta_{\text{Iso}} = 9.6 \times 10^{-5}$, approximately one order of magnitude below Planck's constraint on fully anti-correlated isocurvature perturbations, $\beta_{\text{Iso}}^{\text{Planck}} < 8 \times 10^{-4}$ \cite{Akrami:2018odb}. 

The isocurvature spectral index is generically different from the adiabatic one, and in our case we obtain:
\begin{equation}
n_I =  n_s + \frac{36(6\epsilon_{\phi_*} -2 \eta_{\phi_*})Q_*}{(-3+Q_*)(3+5Q_*)^2} + \frac{12\sigma_{\phi_*} + 4(-3 \epsilon_{\phi_*} + \eta_{\phi_*} + \sigma_{\phi_*})Q_*}{(1+Q_*)(3+5Q_*)}.
\end{equation}
Nevertheless, for our illustrative example, this yields $n_I \approx 0.963$, which differs from $n_s$ by only $\simeq 0.4$\%.

In addition to isocurvature modes, the inflaton also contributes to the effective number of relativistic degrees of freedom while in the dark radiation phase. In this case, the contribution from the background condensate is $\Delta N_{\text{eff}}^\phi =  4.4 (\rho_{\phi}/\rho_{\gamma})$ which is typically quite suppressed, e.g.~$\Delta N_{\text{eff}}^\phi \simeq 9 \times 10^{-5}$ for the parameters employed in figures~\ref{inflationary_dynamics} and \ref{cosmological_dynamics}. More significant is the contribution of decoupled inflaton particles, $\Delta N_{\text{eff}}^{\delta\phi}~=~(4/7) \left(43/4g_*\right)^\frac{4}{3}~\approx~0.027$,  which may be probed with the next generation CMB experiments and large-scale structure surveys \cite{Baumann:2018}. An interesting aspect to emphasize is the fact that, while both these contributions are present during nucleosynthesis, the background field already behaves as dark matter at recombination, while the decoupled inflaton particles typically become non-relativistic around this time or slightly afterwards. Hence, our setup generically predicts a distinct effective number of relativistic species as inferred from light nuclear element abundances and the CMB spectrum, which is a quite distinctive signature.

We note that in the WLI scenario analyzed in \cite{Rosa:2018iff}, where the possibility of inflaton dark matter in warm inflation was first identified, corresponding to the case where only one of the couplings $g_{1,2}$ was non-zero, the inflaton-to-radiation ratio in the post-inflationary dark radiation phase was larger than in the more general case analyzed in this work, where $g_1\neq g_2\neq 0$, yielding larger values for $\Delta N_{\text{eff}}^\phi$. This is essentially due to the smaller values of the dissipation coefficient in the case where only one of the couplings in non-zero, which trigger earlier inflaton oscillations in the quartic potential and thus with a larger amplitude.


\section{Neutrino masses}\label{sec:NuMasses}

As we have showed in the previous section, the Yukawa terms of \eqref{YukLag} lead to dissipative dynamics during an inflationary period with $T_{\text{inf}} \sim 10^{15}$ GeV. These terms also lead to the generation of neutrinos masses at low energies through the seesaw mechanism (see section~\ref{sec:BasicIngredients}): in this model, the high-energy and low-energy physics are related because both dissipation and the seesaw mechanism are realized through the same operators. 

Recalling Eq.~\eqref{eq-seesaw} and Table \ref{Tab:content}, and taking, without loss of generality, $M_R$ and the charged lepton mass matrices to be diagonal, only the Dirac mass matrix $m_D$ remains to be determined. This matrix will be shaped by the $\mathbb{Z}_3$ charge assignments of the left-handed leptons, and the interchange symmetry (the latter imposing that the first and second rows must be identical). Recalling the two possible $\mathbb{Z}_3$ configurations in Table \ref{Tab:content}, we have the following possible structures for the Dirac mass matrix: 
\begin{enumerate}
\item Option I:
\begin{equation}
m^I_D=v\begin{pmatrix} a & b & 0 \\ a & b & 0 \\ 0 & 0 & c \end{pmatrix}, 
\end{equation}
resulting in the following light neutrino masses:
\begin{equation}\label{MassClassI}
m_\nu^{(I)} = \begin{pmatrix} 0, & v \frac{\lvert c\rvert^2}{M_3}, v\frac{(\lvert a \rvert^2 +\lvert b \rvert^2)(M_1+M_2)}{M_1M_2} \end{pmatrix},
\end{equation}
where $\lvert a \rvert^2 +\lvert b \rvert^2=y^2$.

\item Option II:
\begin{equation}
m_D^{(II)}=v\begin{pmatrix} a & b & c \\ a & b & c \end{pmatrix}~, 
\end{equation}
which features two zero eigenvalues:
\begin{equation}\label{MassClassII}
m^{II}_\nu = \begin{pmatrix} 0, & 0, & v\frac{M_1 + M_2}{M_1 M_2} (\lvert a\rvert^2+\lvert b^2\rvert+\lvert c\rvert^2) \end{pmatrix}~,
\end{equation}
where, in this case, $\lvert a\rvert^2+\lvert b^2\rvert+\lvert c\rvert^2=y^2$. Although a spectrum with two massless neutrinos is experimentally ruled out, the separation between the solar and atmospheric neutrino mass-squared differences indicates that two of the neutrinos have relatively similar masses. One may then envisage scenarios where small deviations from the above $\mathbb{Z}_3$-symmetric structure are responsible for the necessary mass-splitting.

\end{enumerate}
Regardless of the charge assignment, given the common prediction of a massless neutrino, it is ultimately possible to identify the effective Yukawa coupling used in the dynamics of warm inflation as being proportional to the mass scale $\sqrt{\Delta m^2_\text{sol}} \sim 0.05$ eV. There is, therefore, a very close relation between inflationary dynamics and low-energy phenomenology, one of the attractive features of this model.

In order to obtain a more realistic mass spectrum, one has to account for the dynamical nature of the Yukawa couplings caused by the large separation between the right-handed neutrino mass scale ($\sim 10^{15}$ GeV) and the electroweak scale ($\sim 10^2$ GeV), through the use of the renormalization group equations (RGEs). In particular, we are interested in the evolution of effective Yukawa coupling $y$, as this simultaneously determines the amount of dissipation during inflation and the scale of the light neutrino masses.

\begin{figure}[h!]
	\centering\includegraphics[width=0.6\textwidth]{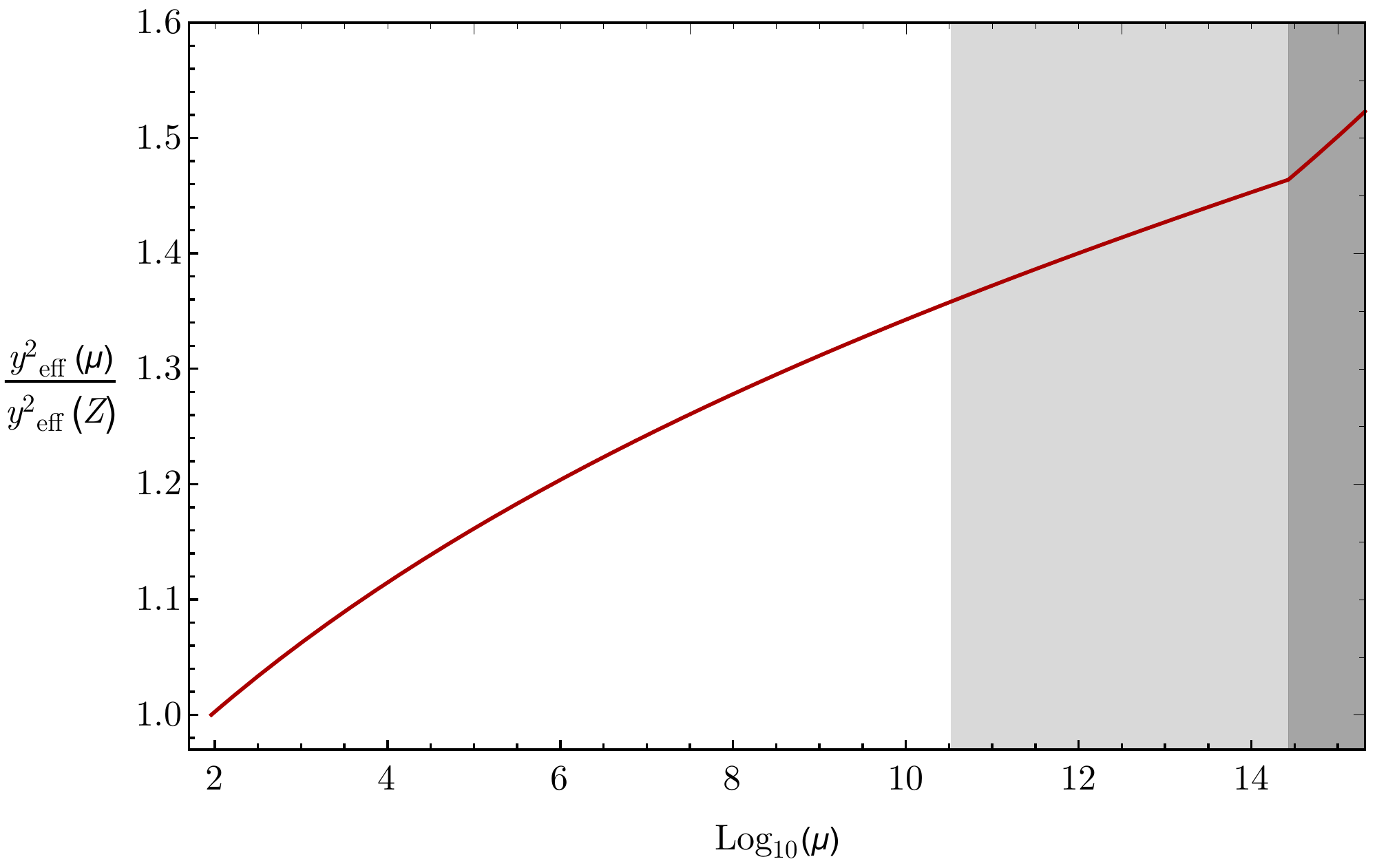} 
	\caption{Running of the squared effective Yukawa coupling  $y^2$ from the $Z$ mass scale to the U(1) symmetry breaking scale $M$. The mass threshold of the lightest RHN ($N_3$) is shown in light gray,  while the mass threshold of the heaviest degenerate RHN ($N_{1,2}$) is shown in dark gray.}
	\label{fig:running}
\end{figure}

For this, we use the \textit{REAP} \textit{Mathematica} package \cite{Antusch:2005gp}, which solves the RGEs of the leptonic sector for the Standard Model plus right-handed neutrinos (SM+RHN) model\footnote{This is well justified because the energy scales of interest here are below the SSB scale of $U(1)_X$ gauge symmetry group: the theory is an effective SM+RHN.}. One finds that $y^2(\mu \sim 10^{15} \text{ GeV})/y^2(\mu \sim 10^{2} \text{ GeV}) \sim 1.4-1.5$, as explicitly shown in figure~\ref{fig:running} for the ``$\mathbb{Z}_3$-II" charge assignment. To obtain the correct neutrino masses at the low-energy scale, one has to account for this effect. 
Let us take the results obtained in section \ref{sec:InflationaryDynamics}, and focus on the example of figures~\ref{inflationary_dynamics} and \ref{inflationary_spectrum}, with $y=0.74,~g_1=0.21, g_2 = 0.17,~M \approx 3.3 \times 10^{15}$ GeV. Using Eq.~\eqref{MassClassII}, it becomes clear that the required Yukawa couplings obtained in section \ref{sec:InflationaryDynamics} already result in masses compatible with the low energy physics:
\begin{equation*}
m^{(\mu=M)}_3 = 2 \frac{y^2 v^2}{M_1} \sim 0.07 \text{ eV} \xrightarrow{\text{running}} m_3^{(\mu=m_Z)} \sim 0.05 \text{ eV}.
\end{equation*}
A more detailed account of RGEs and the running of parameters is outside the scope of this paper, and the interested reader is pointed to, e.g. \cite{Antusch:2003kp, Antusch:2005gp}.

\begin{figure}[h!]
	\centering\includegraphics[width=.6\textwidth]{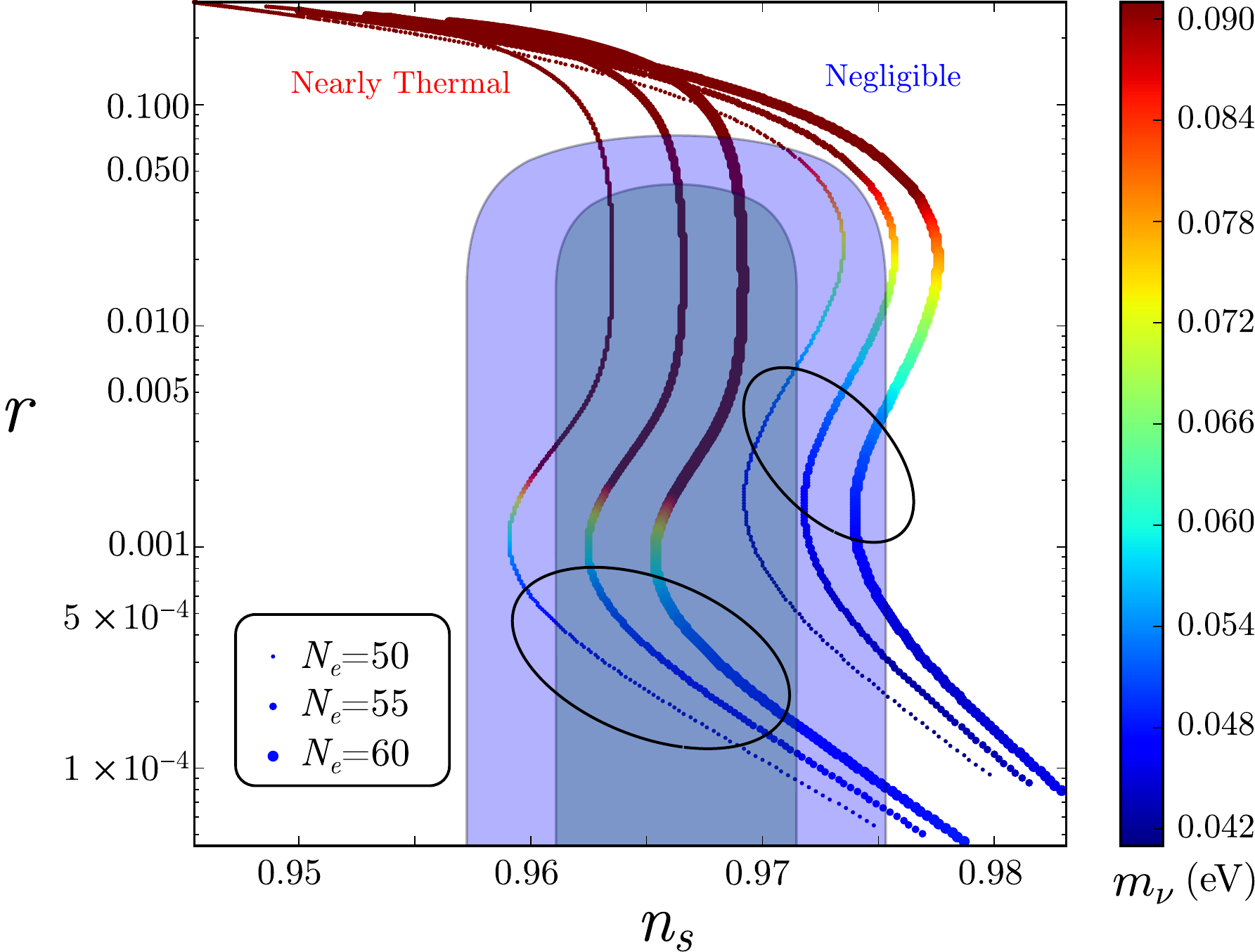}
	\caption{Predictions for inflationary observables ($n_s,~r$) and the largest left-handed neutrino mass eigenstate, for $V(\phi)=\lambda \phi^4$, considering the two limiting cases where inflaton particles are nearly thermalized or have negligible occupation numbers at horizon-crossing of the relevant CMB scales. The blue contours correspond to the Planck TT,TE,EE+lowE+lensing +BK14+BAO 68\% and 95\% C.L. regions \cite{Akrami:2018odb}. The ellipses highlight the parametric regions where the neutrino mass is in agreement with the experimental value.} \label{moneyplotz}
\end{figure}

We can relate inflationary observables and neutrino masses within our setup in the following way. For each value of the dissipative ratio at horizon-crossing, $Q_*$, the number of e-folds of inflation and the amplitude of the primordial curvature spectrum can be used to infer $\phi_*$ and $T_*/H_*$. In turn, these values can be used to determine the combination $(g_1^2+g_2^2)\cos^2\delta/y^2$ that yields the magnitude of the dissipation coefficient. We set the effective Yukawa coupling $y$ at the minimum value for which the equilibrium condition $\Gamma_N>H$ is satisfied at horizon-crossing, and the mass scale $M$ can be set by demanding that $T\gtrsim M_{1,2}$ throughout inflation\footnote{More specifically, we find numerically that requiring $T_*>1.7M_1$ is enough to ensure that the right-handed neutrinos remain sufficiently light for $\sim$60 e-folds of inflation after horizon-crossing of CMB scales.}. In this analysis we set $g_2=0.8g_1$ for concreteness, noting that similar results are obtained if there is no large hierarchy between these couplings. In figure~\ref{moneyplotz} we show the results of this analysis, simultaneously showing the predictions for the inflationary observables and the mass $m_3$ of the heaviest left-handed neutrino in the ``$\mathbb{Z}_3$-II" charge assignment (which is independent of the mass of the third right-handed neutrino, $M_3$).

As one can see in this figure, agreement with CMB observations corresponds to a heaviest neutrino mass $0.04\ \mathrm{eV} <m_3<0.1\ \mathrm{eV}$, which is quite remarkable. The most realistic scenarios, with $m_3\simeq 0.05$ eV, correspond to dissipative ratios at horizon-crossing $0.1\lesssim Q_* \lesssim 1$.

Neither of the above charge assignments can, however, account for a realistic neutrino mixing pattern compatible with neutrino oscillation experiments, requiring additional degrees of freedom involved in the seesaw mechanism. By including these, we should be careful in not spoiling the already successful realization of warm inflation with a realistic mass scale for the heaviest left-handed neutrino. 

One possible option is to consider an $SU(2)_L$ scalar triplet $\Delta_L$. The vacuum expectation value of this field generates a renormalizable Majorana mass term for the light neutrinos through the type II seesaw mechanism \cite{Schechter:1980gr}. This would then result in a mixed type I+II seesaw, with $m_\nu \rightarrow m_{LL} - m^T_D M^{-1}_{R} m_D$. These sort of scalars are common in UV-completions of the SM, such as LRSM \cite{Mohapatra:1974gc} and $SO(10)$ GUTs \cite{Lazarides:1980nt, Hambye:2003ka}. The downside of this approach is the introduction of a large number of new parameters in the theory.

A more economical path, on which we will focus our discussion, is the introduction of a flavour-symmetry breaking ``flavon" field, $\chi$, singlet under the gauge group but with a non-trivial $\mathbb{Z}_3$-charge. For instance, choosing a charge $\omega^2$ for the flavon field allows for a dimension-5 term $\chi \overline{N}_3 H L_L/\Lambda$, where $\Lambda\gtrsim M$ is a heavy mass scale depending on the UV-completion of the model. Upon acquiring an expectation value $\langle \chi \rangle\equiv v_\chi \ll \Lambda$, the flavon field therefore allows for small Yukawa couplings involving the third right-handed neutrino, which were thus far forbidden by the $\mathbb{Z}_3$ symmetry.

Since the $\mathbb{Z}_3$-II charge assignment yields a simpler neutrino mass spectrum, depending only on a single effective Yukawa coupling in the $\mathbb{Z}_3$-symmetric limit, we will henceforth consider only this option. The non-zero flavon VEV then allows for a Dirac mass matrix of the form:
\begin{equation}\label{mDpert}
m_D = \begin{pmatrix} a & b & c \\ a & b & c  \\0 & 0 &0 \end{pmatrix} \xrightarrow{\chi \rightarrow v_\chi} \begin{pmatrix} a & b & c \\ a & b & c  \\d & e & f \end{pmatrix}.
\end{equation}
The interchange symmetry is unaffected by these terms, and consequently the setup still predicts one massless neutrino. The new entries $(d,e,f)$ in \eqref{mDpert} in the Dirac mass matrix are $\mathcal{O}(v_\chi /\Lambda)$ and can therefore be treated as small deviations, yielding a light neutrino mass matrix of the form:
\begin{equation}\label{mnu2}
m_\nu = -v^2 \begin{pmatrix} \frac{2 a^2}{M_1}+\frac{d^2}{M_3} &  \frac{2 a b}{M_1}+\frac{d e}{M_3} &  \frac{2 a c}{M_1}+\frac{d f}{M_3}  \\ . & \frac{2 b^2}{M_1}+\frac{y^2}{M_3} &  \frac{2 b c }{M_1}+\frac{e f}{M_3} \\ . & . &  \frac{2 c^2}{M_1}+\frac{f^2}{M_3} \end{pmatrix}.
\end{equation}
Note that $M_3$ is, so far, a free parameter, therefore a suppression of $(d,e,f) \propto v_\chi/\Lambda$ affecting the flavon terms can be countered by a lower $M_3$. Therefore, a hierarchy between $M_1$ and $M_3$, $M_1 \gg M_3$, may lead to sizable flavonic contributions, e.g. $(2a^2/M_1) \gtrsim (d^2/M_3)$, to the effective neutrino mass matrix.

As a simple example, let us consider the case $d,e=0$, which leads to:
\begin{equation}
\begin{aligned}\label{approxmasses}
m_2\simeq  f^2\frac{v^2}{M_3} \frac{ a^2 + b^2}{y^2}, \quad m_3 \simeq \frac{2 y^2 v^2}{M_1} + f^2\frac{v^2}{M_3} \frac{c^2}{y^2},
\end{aligned}
\end{equation}
where we expanded the non-trivial eigenvalues of $m_\nu m_\nu^\dagger$ to second-order in $f$ and, for simplicity, took all couplings to be real. This shows that a single flavonic term in the Dirac Yukawa matrix is is sufficient to lift the unrealistic degeneracy of the neutrino mass spectrum. Inspecting Eq.~\eqref{approxmasses}, it is clear that the choice $c=0$ protects $m_3$ against unwanted contributions. In this limit, a realistic value for the second non-vanishing neutrino mass $m_2$ can be obtained for:
\begin{equation}
f^2 \approx 4 \times 10^{-6} \frac{M_3}{10^{10}\text{ GeV}}~.
\end{equation}
where we have included the Yukawa running. This illustrates that the third right-handed neutrino, which we recall does not interact directly with the inflaton field, can be much lighter than the other two, noting that the flavon-induced Yukawa couplings can be suppressed for $v_\chi\ll \Lambda$. 

We must ensure that the introduction of an additional flavon field and its induced $N_3$ Yukawa couplings do not spoil the successful realization of warm inflation obtained earlier with only two right-handed neutrinos. If, on the one hand, the flavon field acquires its VEV during inflation, this may lead to a direct coupling between the inflaton and the third right-handed neutrino through dimension-5 operators of the form $g_3\overline{N_3^c} \Phi_{1,2} N_3$, where $g_3=\mathcal{O}(v_\chi/\Lambda)$. These reintroduce dangerous thermal corrections to the inflaton mass that could spoil inflation if $N_3$ is thermalized, unless $g_3\ll 1$. Given that $N_3$ can be thermalized via decays and inverse decays, with $\Gamma_{N_3}/H \sim |y_3|^2 (T/H)$ during inflation, with $|y_3|^2 \equiv |d|^2 + |e|^2 + |f|^2$, this leads to an upper bound on this effective coupling $|y_3|^2 \lesssim 10^{-5}$. 

On the other hand, it is possible that the flavon field only acquires a VEV after inflation ends, in which case $N_3$ is not thermally excited during inflation and therefore there are no additional thermal corrections to the inflaton's mass. This would require $m_\chi \gtrsim H\gtrsim v_\chi$ during inflation. This could be due to e.g.-~Planck-suppressed non-renormalizable contributions to the flavon potential or a non-minimal coupling to gravity, $\xi \chi^2 R$, which easily induce $\mathcal{O}(H)$ masses to scalar fields unprotected by any symmetry. In the case of a non-minimal coupling to gravity, the induced flavon mass becomes highly suppressed once radiation becomes dominant ($R\simeq 0$), so that the spontaneous breaking of the $\mathbb{Z}_3$ symmetry by the flavon field would naturally be triggered once inflation ends. One may worry that this could lead to the formation of domain-walls that would eventually overclose the Universe, but we note that the spontaneous breaking of the U(1) symmetry via the $\Phi_i$ VEVs already leads to soft $\mathbb{Z}_3$-breaking terms such as $\chi^2\Phi_1 \Phi_3^\dagger \xrightarrow{SSB}  M M' \chi^2$ that ensure that the flavon potential has a single global minimum. In this scenario $v_\chi\lesssim 10^{12}-10^{13}$ GeV, such that if the scale of the non-renormalizable flavon operators $\Lambda\gtrsim M\sim 10^{15}$ GeV, the $N_3$ Yukawa couplings should also be suppressed in this case, $y_3\lesssim 10^{-3}-10^{-2}$.

We thus conclude that, in both scenarios, the third right-handed neutrino must be relatively light to accommodate a realistic neutrino mass spectrum, $M_3\lesssim 10^{10}-10^{11}$ GeV.


\section{Leptogenesis} \label{sec:Leptogen}

It is well known that the addition of right-handed neutrinos provides an elegant mechanism for generating the cosmological baryon asymmetry known as leptogenesis \cite{Fukugita:2003en}, where lepton number violation is a consequence of the Majorana mass terms and C/CP-violation arises through the neutrino Yukawa couplings. In its simplest realization, a lepton asymmetry is then generated via the out-of-equilibrium decays of an initially thermal population of right-handed neutrinos. For hierarchical Majorana masses, the lightest right-handed neutrino is responsible for producing the largest contribution to the lepton asymmetry, which is later converted into a baryon asymmetry by electroweak sphaleron processes, which violate $B+L$ while preserving the combination $B-L$ \cite{Hirsch:2001dg}.

It is then clear that leptogenesis can naturally be incorporated within our setup, where the underlying U(1) gauge symmetry can be identified with lepton number or $B-L$, two of the right-handed neutrinos are degenerate (at the minimum of the inflaton potential) and the third right-handed neutrino is typically much lighter in order to obtain a realistic light neutrino mass spectrum. 

As we have seen in the previous section, consistency of warm inflation requires $N_3$ to only be thermally produced after inflation, either because its Yukawa couplings are suppressed by the ratio $v_\chi/\Lambda\ll 1$ or because the flavon field only breaks the $\mathbb{Z}_3$ family symmetry after inflation ends. In either case, a thermal population of $N_3$ is produced in the radiation bath after inflation if the effective $N_3$ Yukawa coupling $y_3$ is not too suppressed.

The CP asymmetry due to $N_3$ decays is given by \cite{Davidson:2008bu}
\begin{equation}\label{epsparam}
\begin{aligned}
\varepsilon &= \frac{\Gamma (N_3 \rightarrow L_j + H) - \Gamma (N^\dagger_3 \rightarrow L^\dagger_j + H^\dagger)}{\Gamma (N_3 \rightarrow L_j + H) + \Gamma (N^\dagger_3 \rightarrow L^\dagger_j + H^\dagger)}  \\
&= \frac{1}{8 \pi (Y^\dagger_\nu Y_\nu)_{33}} \sum_{i \neq 3} \text{Im} \left( \left[(Y_\nu^\dagger Y_\nu)_{3i} \right]^2\right)  \left( f \left(\frac{M_i^2}{M_3^2}\right)+g \left(\frac{M_i^2}{M_3^2}\right) \right), 
\end{aligned}
\end{equation}
where 
\begin{equation}
f(x) = \sqrt{x}\left[ 1-(1+x)\ln\left(\frac{1+x}{x}\right)\right], \quad g(x)=\frac{\sqrt{x}}{1-x},
\end{equation}
and $f(x)$ results from the interference between the tree-level decay amplitude and the 1-loop vertex correction, whereas $g(x)$ stems from the absorptive part of the neutrino self-energy\footnote{Note that the expression \eqref{epsparam} differs from the one presented in \cite{Hirsch:2001dg} due to a different convention in the definition of $Y_\nu$, which are related through a $\mathrm{h.c.}$ transformation.}.

This CP asymmetry of the $N_3$ decays leads to a lepton asymmetry, diluted by wash-out processes, such as scattering and inverse decays. The final lepton-to-entropy ratio is then given by:
\begin{equation}
Y_L= d \frac{\epsilon}{g_*},
\end{equation}
where $g_*$ is the effective number of degrees of freedom, and $d$ is the dilution factor, which is well approximated by \cite{Kolb:1983ni, Nielsen:2001fy}:
\begin{equation}
d=\frac{1}{2\sqrt{K^2+9}}, \qquad K=\left.{\Gamma_{N_3}\over 2H}\right|_{T=m_3}\simeq \frac{M_P}{1.7 \times 8 \pi \sqrt{g_*}}\frac{(Y_\nu^\dagger Y_\nu)_{33}}{M_3},
\end{equation}
for $K \lesssim 10$, which is the region of interest in the present case. 

Finally, the leptonic asymmetry is converted into a baryonic asymmetry through sphaleron processes, yielding a baryon-to-entropy ratio:
\begin{equation}
Y_B  = \frac{C}{C-1} Y_L, \qquad C = \frac{8N_F +4N_H}{22N_F+13N_H},
\end{equation}
where $N_F$ is the number of families and $N_H$ the number of Higgs doublets. In this model, which after inflation becomes simply the SM with right-handed neutrinos, $N_F=3$ and $N_H=1$, giving $C\simeq1/3$. The baryon asymmetry can be experimentally measured through the BBN and CMB,  with $Y_B \simeq (8-10) \times 10^{-11}$ \cite{Cyburt}.

\begin{figure}[htbp]
	\centering\includegraphics[width=0.6\textwidth]{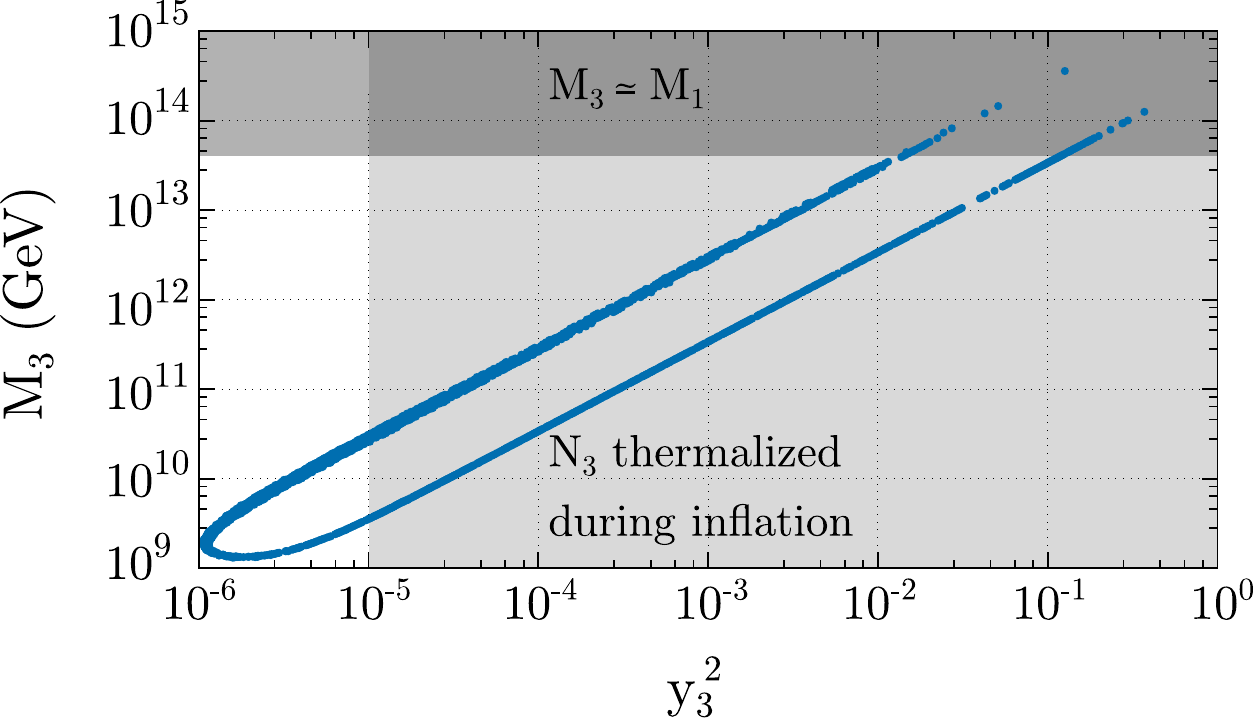}
	\caption{Mass of the lightest right-handed neutrino, $N_3$, as a function of its effective Yukawa coupling, $|y_3|$. All neutrino observables are required to be within the 3$\sigma$ experimental range and we have fixed $y=0.74$, $M_1=4.6 \times 10^{14}$ GeV, and $Y_B=7.5\times 10^{-11}$. The light-gray shaded region shows the forbidden region if the $N_3$ neutrino thermalizes during inflation, and the gray shaded region estimates the breakdown of the approximation $M_3\ll M_1$.}\label{fig:M3}
\end{figure}

We have performed a numerical parameter scan of our model, matching the full neutrino mass matrix to the measured values of the neutrino mass differences and mixing parameters within three standard deviations \cite{deSalas:2017kay}, fixing the effective Yukawa coupling $y$ and the heaviest neutrino masses $M_1=M_2$ to values yielding a technically and observationally consistent warm inflation scenario. Requiring that the baryon symmetry resulting from out-of-equilibrium $N_3$ decays matches the observational value, it is then possible to relate the mass of the lightest right-handed neutrino, $M_3$ (unconstrained by inflation), with its effective Yukawa coupling $y_3$ defined above. The results of this analysis are shown in figure~\ref{fig:M3}.

Taking into account our previous discussion on the size of the $N_3$ Yukawa couplings, we see that a successful model consistently yielding inflation, a realistic neutrino mass spectrum and the observed cosmological baryon asymmetry requires $M_3\sim 10^{9}-10^{11}$ GeV, consistently with the Davidson-Ibarra bound \cite{Davidson:2002qv}, with suppressed Yukawa coupling $|y_3|\sim 10^{-3}-10^{-2}$ as expected from their flavon-induced origin.



\section{Summary and future prospects} \label{sec:Summary}

The present work addresses the main shortcomings of the SM except those involving fine-tuning issues (which arguably are not true inconsistencies of the theory), within a unified model: 1) inflation is driven by a slowly rolling scalar field, 2) neutrino masses are generated via the seesaw mechanism, with right-handed neutrinos being the only particles interacting directly with the inflaton field; 3) the remnant inflaton field naturally accounting for dark matter at late times; and 4) a cosmological baryon asymmetry is generated via thermal leptogenesis. This is implemented within the first realization of warm inflation within a concrete extension of the SM, based on the generic Warm Little Inflaton scenario. This is a minimal extension with the addition of three right-handed neutrinos and three complex scalar fields that are singlets under the SM gauge group but charged under a gauged U(1) symmetry identified with lepton number or $B-L$. Moreover, all these problems are addressed without modifying the gravitational sector, making it less sensitive to the UV-completion of the model.

The model is endowed with two additional discrete symmetries: an interchange symmetry under which $N_1\leftrightarrow N_2$ and $\Phi_1\leftrightarrow i\Phi_2$, leaving $N_3$ and $\Phi_3$ unchanged, and a $\mathbb{Z}_3$ family symmetry. The inflaton field is identified with the relative phase of the complex fields $\Phi_1$ and $\Phi_2$, and interacts directly only with the right-handed neutrinos $N_1$ and $N_2$. These interactions, alongside the decay of the right-handed neutrinos into leptons and Higgs degrees of freedom, are responsible for dissipative effects that sustain a warm thermal bath during inflation, with the interchange symmetry protecting the scalar potential against large thermal corrections that could prevent a slow-roll evolution. We have shown that such a scenario consistently allows for a sufficiently long inflationary period with a quartic scalar potential, with a scalar spectral index and tensor-to-scalar ratio within the range allowed by Planck data, as in the original WLI model. In addition, the strong dissipation regime is attained towards the end of inflation, causing the radiation (i.e.~right-handed neutrinos, leptons and Higgs doublets) to dominate smoothly at the end of the slow-roll regime. At this stage, the ratio $T/H$ also becomes sufficiently large to allow for the thermal production of the remaining SM degrees of freedom.

One of the most interesting features of this setup is that the scale of the right-handed neutrino masses, $M_{1,2}\sim 10^{14}-10^{15}$ GeV, and effective Yukawa couplings, $y\lesssim 1$, is fixed by the constraints of realizing warm inflation near thermal equilibrium at temperatures above the right-handed neutrino mass threshold, since otherwise dissipative effects would be exponentially suppressed. This automatically yields the scale of the heaviest left-handed neutrino mass, via the seesaw mechanism, in agreement with neutrino oscillation experiments, $m_3\simeq 0.05$ eV (taking into account the small running of the Yukawa couplings). The interchange symmetry also leads to a specific prediction of at least one massless neutrino.

A full agreement with experimental results for neutrino masses and mixings requires an extension of this basic setup with e.g.~a flavon field that breaks the $\mathbb{Z}_3$ family symmetry, since otherwise the third-right handed neutrino is not involved in the seesaw mechanism. We have shown that this can be done without spoiling the successful realization of warm inflation, since $N_3$ need not be thermally excited during inflation.

Our setup also offers natural solutions to two other important cosmological problems. First, the interchange symmetry makes the inflaton stable at late times and protects its mass from large radiative corrections. Since the right-handed neutrino masses are bounded, $N_{1,2}$ are only relativistic and thermally produced during inflation, decoupling from the dynamics once the temperature drops below their mass threshold at the end of inflation. The resulting inflaton remnant thus corresponds to a cold and weakly interacting fluid that may naturally account for dark matter for inflaton masses $\lesssim$ 1 eV. Second, a thermal population of the lightest right-handed neutrino is necessarily generated after inflation, and its out-of-equilibrium decays can produce a lepton asymmetry that is later converted into the observed cosmological baryon asymmetry by electroweak sphaleron processes. In our setup, the three right-handed neutrinos play separate roles, with $N_1$ and $N_2$ being responsible for warm inflation and $N_3$ for thermal leptogenesis after warm inflation, but all within the same framework yielding a realistic pattern for neutrino masses and mixings.

In addition, the model can be probed via a plethora of different observables, with both low-energy neutrino experiments and astrophysical observations, as best exemplified by figure~\ref{moneyplotz}.  Warm inflation leads to a specific consistency relation between the tensor spectral index and the tensor-to-scalar ratio \cite{Cai:2010wt, Bartrum:2013fia}, and within our particular setup we predict the latter to lie in the range $r\sim 10^{-4}-10^{-3}$ given the measured scale of neutrino masses. Other observables include non-Gaussian features with a particular ``warm" shape that may be accessible in the near future and cold dark matter isocurvature modes roughly an order of magnitude below the current Planck sensitivity, and anti-correlated with the main adiabatic curvature perturbations. The fact that the inflaton remnant behaves as dark radiation during BBN and cold dark matter at recombination also yields a very distinctive probe of our scenario. In conjunction, all these observables make our setup quite distinguishable from other interesting attempts to simultaneously address the same problems (e.g. \cite{Ballesteros:2016euj, Ballesteros:2016xej}).

It would be interesting to try to extend the present warm inflation scenario into the full strong dissipation regime, i.e.~$Q_*\gtrsim 100$, where the inflationary dynamics becomes less reliable on the UV completion \cite{Berera:1999ws, Berera:2004vm, Motaharfar:2018zyb, Das:2018rpg}. Within the current understanding of fluctuations in warm inflation, our setup would predict a too blue-tilted power spectrum in this regime, due to the interplay between inflaton and radiation fluctuations. This may require modifying the particle content and the inflaton potential, as recently proposed in the modified WLI-construction of \cite{Bastero-Gil:2019gao}, or a better understanding of the non-equilibrium properties of the thermal bath, which may have a significant impact on observational predictions (e.g.~\cite{BasteroGil:2011xd}). 
 
 A natural possibility to investigate in the future is also embedding the minimalistic setup presented in this work within grande unified theories, particularly those based on $SO(10)$ or larger gauge groups containing the latter as a sub-group, since right-handed neutrinos are automatically included in the fundamental representation of $SO(10)$ alongside the known quarks and leptons. As we mentioned in section~\ref{sec:BasicIngredients}, the first option for the $\mathbb{Z}_3$ charge assignments could accommodate such an embedding, and potentially both the latter and the discrete interchange symmetry can have measurable effects on the low-energy particle spectrum that could further help testing this model.

\acknowledgments{
We thank Mar Bastero-Gil, Arjun Berera and Rudnei Ramos for useful discussions on this topic. This work was supported by the FCT Grant No.~IF/01597/2015.  The work of M.\,L. is funded by Funda\c{c}\~ao para a Ci\^encia e Tecnologia-FCT Grant No.PD/BD/150488/2019, in the framework of the Doctoral Programme IDPASC-PT. ML was initially supported by Grant No. BI/UI97/8364/2019, in the context of the FCT Grant No.~IF/01597/2015. 
J.\,G.\,R. and L.\,B.\,V. are supported by the CFisUC project No.~UID/FIS/04564/2019. J.~G.R. is also supported by the projects PTDC/FIS-OUT/28407/2017 and the  ENGAGE SKA (POCI-01-0145-FEDER-022217). L.B.V is also supported by FCT Grant PD/BD/140917/2019 and partially by the CIDMA project No. UID/MAT/04106/2020.

}

\appendix

\section{Inflaton self-energy}\label{App:InflatonSelf-Energy}
One of the most important features of the WLI paradigm is the removal of the leading contributions to $V(\phi)$ that preclude inflation, as discussed in the main text. An alternative way to see this is by computing the inflaton self-energy generated from the couplings to the $N_{1,2}$ fermions. To do so, one can start with the relevant Lagrangian \eqref{YukLag} after SSB of the $U(1)_X$, working in the RH neutrino mass basis\footnote{Hence the appearance of the phases $\delta_i$ resulting from the appropriate rotations.}:
\begin{equation}\label{InfSElag}
-\mathcal{L}_{\phi, N} = \overline{N^c}_{i} \left( \frac{1}{2} M_i  + \frac{1}{2}V_i e^{i \delta_i} \delta \phi + \frac{1}{4}f_i \delta \phi  \delta \phi  \right) N_{i} + h.c. ,
\end{equation}
where summation over the two RH neutrinos is implied and the background value of the inflaton field was separated from its fluctuations, by replacing $\phi \rightarrow \phi + \delta \phi$ and expanding the exponential in powers of $\delta \phi/M$ up to quadratic order
%
\begin{equation}
e^{i\phi/M} = e^{i\phi/M}e^{i\delta\phi/M} 
= e^{i\phi/M}\sum_{n=0}^\infty \frac{(i\delta \phi /M)^n}{n!} = e^{\phi/M} \left( 1 + i\delta \phi/M + \delta \phi \delta \phi/M^2 +...\right).
\end{equation}
Here, $M_i$, $V_i$ and $f_i$ take real values, and are related to each other (see below). One can now define Majorana fermions as
%
\begin{equation}
\label{majoranafermions}
\mathbf{N}_i=( N_i +  N_i^c), \qquad \overline{\mathbf{N}_i}=\overline{( N_i +  N_i^c)}, \qquad N_i = P_R \mathbf{N}_i, \qquad  N_i^c = P_L \mathbf{N}_i, 
\end{equation}
and write \eqref{InfSElag} as 
\begin{equation}
-\mathcal{L}_{\phi, N} = \bigg(\frac{1}{2} M_i \overline{\mathbf{N}_i} \mathbf{N}_i + \frac{1}{2}V_i \overline{\mathbf{N}_i} e^{i \gamma_5 \delta} \delta \phi \mathbf{N}_i + \frac{1}{4}f_i \overline{\mathbf{N}_i}\delta \phi \delta \phi \mathbf{N}_i \bigg), 
\end{equation}
where $f_i = - M_i/M^2$ and $V_i^2 = M_j^2/M^2$, $i=1,2$ and $j\neq i=2,1$, with 
\begin{equation}
	\begin{aligned} \label{Vertices and Phase}
		&M_1^2= \frac{M^2}{4}\left[(g_1+g_2)^2\cos^2(\phi/M)+(g_1-g_2)^2\sin^2(\phi/M)\right],\\
		&M_2^2= \frac{M^2}{4}\left[(g_1-g_2)^2\cos^2(\phi/M)+(g_1+g_2)^2\sin^2(\phi/M)\right], \\
		&-\delta = \tan^{-1}\left(\frac{g_-}{g_+} \cot(\phi/M)\right) + \tan^{-1}\left(\frac{g_- }{g_+} \tan(\phi/M)\right) ,
	\end{aligned}
\end{equation}
%
where $g_\pm$ was defined in \eqref{Masses}. 

\begin{figure}[h!]
	\centering\includegraphics[width=0.6\textwidth]{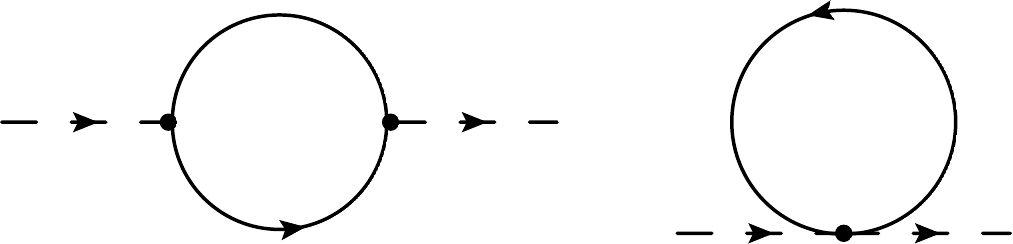} 
	\caption{Feynman diagrams contributing to the inflaton self-energy, \eqref{pi-cubic} and \eqref{pi-tadpoles}, respectively. The dashed line represents $\delta \phi$ while the solid one represents $\mathbf{N}_i$.}
	\label{fig:self-en}
\end{figure}

There are two different contributions to the inflaton self-energy i) from the 4-point interaction $\delta \phi \delta \phi N  N $ terms, ii) from the 3-point Yukawa interaction $\delta \phi N N$ terms, represented in figure~\ref{fig:self-en}. The former is readily computed:
\begin{equation}
\begin{aligned}
\label{pi-tadpoles}
-i \Pi_{\text{4P}} (0) &=  \frac{1}{2}(-1) \loopp \text{Tr}\left[  i f_1 \fproppone + i f_2 \fpropptwo \right] \\
&=\frac{1}{2}\frac{1}{M^2} \loopp \left(\frac{4M_1^2}{p^2-M_1^2}+\frac{4M_2^2}{p^2-M_2^2}\right).
\end{aligned}
\end{equation}
where the symmetry factor of $1/2$ is due to the Majorana nature of $N$ \cite{Gluza:1991wj} and we used the relation between $f_i$ and $M_i$. 


The cubic interactions are given by 
\begin{equation}
-i \Pi_{\text{3P}}^i(0) = - \frac{1}{2}V_i^2 \loopp \frac{\text{Tr}\left[\left((\slashed{p}+M_i)( e^{i \delta} P_R +  e^{-i \delta} P_L)\right)^2\right]}{(p^2-M_i^2)^2},
\end{equation}
where
\begin{equation}
\text{Tr}\left[\left((\slashed{p}+M_i)( e^{i \delta} P_R +  e^{-i \delta} P_L)\right)^2\right] =4 p^2 + 4 (\cos^2\delta - \sin^2\delta)M_i^2 = 4p^2+4\cos(2\delta)M_i^2.
\end{equation}
so that the inflaton self-energy correction given by the Yukawa interaction is
\begin{equation}
-i \Pi_{\text{3P}}^i(0) = - \frac{1}{2} V_i^2 \loopp \left( \frac{4p^2+4 \cos(2\delta)M_i^2}{(p^2-M_i^2)^2}\right),
\end{equation}
Summing over the two fermionic contributions, from $N_1$ and $N_2$,
\begin{equation}
\label{pi-cubic} 
 -i\Pi_{\text{3P}}(0) = 
-\frac{1}{2}\frac{1}{M^2} \loopp \bigg(\frac{4M_2^2}{(p^2-M_1^2)} + \frac{4M_1^2}{(p^2-M_2^2)} + \frac{8 M_1^2 M_2^2\cos^2\delta}{(p^2-M_1^2)^2} + \frac{8 M_1^2 M_2^2\cos^2\delta}{(p^2-M_2^2)^2} \bigg).
\end{equation}
By combining \eqref{pi-tadpoles} with \eqref{pi-cubic}, the leading order contributions (in $p$) to the inflaton self-energy cancel. As a result, there are no quadratically divergent terms: only log-divergent terms and finite terms. This holds true in the high temperature limit (HTL), where the masses in the propagators can be neglected when compared to the magnitude of $M_i~\ll~p~\sim~T$.



\section{Fermion decay width and thermal masses}\label{AppB:FDW}

\subsection*{Right-handed (RH) neutrino Decay Width}

Knowledge of the RH neutrino decay width is required for finding the dissipation coefficient, essential for the WI inflationary dynamics. The relevant Lagrangian after the $SU(2)$ contractions is:
\begin{equation}
	\mathcal{L}_{\text{Yuk}, \text{lep}} =   - Y( \overline{\mathbf{N}} P_L h_0 \nu_L - \overline{\mathbf{N}}P_L h^+ l^{\pm}_L) +h.c. , \label{YukawaL2}
\end{equation}
following the notation of \eqref{majoranafermions}. Above the EWSB scale, both charged and neutral processes contribute to the RH neutrino decay width in a similar fashion, as they are mediated by the same Yukawa couplings. Furthermore, the $h.c.$ term above can be recast as $\overline{\nu_L} h_0^* P_R \mathbf{N} \rightarrow \overline{\mathbf{N}} P_R \nu_L^c h_0^*$, by making use of $-\mathbf{N}^T C^{-1} = \overline{\mathbf{N}}$ \cite{Gluza:1991wj}, and $\gamma^0 C \psi^* = \psi^c$\cite{Pal:2010ih}, resulting in four different processes, all of which identical due to the nature of the Cutkosky rules \cite{Cutkosky:1960sp}.

Following the finite-temperature generalization of the Cutkosky rules  \cite{Kobes:1986wb},
\begin{equation}
\text{Im}\Sigma=(1+e^{-\beta p_0})\hat{\Gamma}_d, \qquad \hat{\Gamma}_d = \frac{1}{2} \sum _z F(y_1, y_2, z),
\end{equation}
where $\Sigma$ and $\hat{\Gamma}_d$ is the self-energy and decay width of the RH neutrinos, respectively. The sum in $z$ is over different possibilities of circling internal vertices, and $y_1$ ($y_2$) is the vertex associated with the incoming (outgoing) momenta, and is uncircled (circled)\footnote{In the Cutkosky rules and their finite-temperature generalization, circled vertices are the $h.c.$ of the uncircled vertices.}. At the 1-loop level, there are no internal vertices
\begin{equation}
\label{eq:Gamma_i}
\text{Im}\Sigma_i(p) =\frac{1}{2}(1+e^{-\beta p_0}) {(Y Y^\dagger)}_{ii} \int \frac{d^4k}{(2\pi)^4} P_R \Delta_s^+(k) P_R \Delta_f^+(p-k), 
\end{equation}
in which no summation over repeated indexes is implied, and \cite{Kobes:1986wb}
\begin{eqnarray}
	\Delta_s^+(p)&=& 2\pi [\theta(p_0)+n_B(p_0)]\delta(p^2-m^2),\\
	\Delta_f^+(p)&=& 2\pi [\theta(p_0)-n_F(p_0)](\slashed{p}-m)\delta(p^2-m^2).
\end{eqnarray}
In the massless limit, the above product is equal to
\begin{subequations}
	\begin{align}
 \Delta_s^+(k) \Delta_f^+(p-k) =&(2\pi)^2 (\slashed{p}-\slashed{k}) \delta((p-k)^2) \delta (k^2) \theta(p_0-k_0) \theta(k_0) \label{first-term} \\
		&+(2\pi)^2 (\slashed{p}-\slashed{k}) \delta((p-k)^2) \delta (k^2)\theta(p_0-k_0) n_B(k_0) \label{second-term}\\ 
		&-(2\pi)^2 (\slashed{p}-\slashed{k}) \delta((p-k)^2) \delta (k^2) \theta(k_0) n_F(p_0-k_0) \label{third-term}\\
		&-(2\pi)^2 (\slashed{p}-\slashed{k}) \delta((p-k)^2) \delta (k^2) n_B(k_0)n_F(p_0-k_0). \label{fourth-term}
	\end{align}
\end{subequations}
For \eqref{first-term}, using 
%
\begin{equation}
	\int d^4k \theta(k_0) \delta(k^2) =  \int d^3k \frac{1}{2 k_0},  \quad \text{and} \quad \int d^4p \delta((p-k)^2) = \int d^4p d^4k' \delta(k'^2) \delta((p-k)-k'), 
\end{equation}
one obtains
\begin{equation}
	\begin{aligned}
		&\loopk (2\pi)^2 (P_R)^2 (\slashed{p}-\slashed{k})\theta(p_0-k_0)\theta(k_0)\delta(k^2)\delta((p-k)^2)\\
		&= (2\pi) \int \frac{d^3k }{(2\pi)^3} P_R \frac{\slashed{p}-\slashed{k}}{4\mathbf{|k|}\mathbf{|p-k|}} \delta(p_0-\mathbf{|k|}-\mathbf{|p-k|}) \label{first-term-result}.
	\end{aligned}
\end{equation}
Dropping the $\delta(k_0+\mathbf{|k|})$ contributions (associated with scatterings and other processes \cite{Laine:2016hma}) and keeping in mind that
	\begin{equation}
		n_B(k_0)n_F(p_0-k_0) = (1+e^{\beta p_0})^{-1}(1+n_B(k_0)-n_F(p_0 - k_0)),
	\end{equation}
we find the results
	\begin{eqnarray}
		&&\eqref{second-term} = \eqref{first-term-result}*n_B(\mathbf{|k|}), \nonumber \\
		&&\eqref{third-term} = -\eqref{first-term-result}*n_F(\mathbf{|p-k|}), \\
		&&\eqref{fourth-term} =  -(1+e^{\beta p_0})^{-1} \big(\eqref{first-term}+\eqref{second-term}+\eqref{third-term}\big). \nonumber
	\end{eqnarray}
Therefore, the full contribution to the loop is
%
\begin{equation}
\left(1-(1+e^{\beta p_0})^{-1}\right) \,2\pi\, P_R \,
\int \frac{d^3k}{(2\pi)^3} (1+n_B(\mathbf{|k|})-n_F(\mathbf{|p-k|})) \delta(p_0-\mathbf{|k|}-\mathbf{|p-k|}).
\end{equation}
Recalling \eqref{eq:Gamma_i} and using $(1+e^{-\beta p_0})(1-(1+e^{\beta p_0})^{-1})=1$,
%
\begin{equation}
	\begin{aligned}
		&\text{Im}\Sigma_i(p) =\\ &\frac{1}{2}{(Y Y^\dagger)}_{ii} (2\pi) \int \frac{d^3k}{(2\pi)^3} (P_R) \frac{\slashed{p}-\slashed{k}}{4 \mathbf{|k|} \mathbf{|p-k|}} (1+n_B(\mathbf{|k|})-n_F(\mathbf{|p-k|}))\delta(p_0-\mathbf{|k|}-\mathbf{|p-k|}). \label{Intermediate Decay Width}
	\end{aligned}
\end{equation}
Taking the limit where $p$ and $k$ are aligned,
\begin{equation}
	\slashed{p}-\slashed{k} = \slashed{p}\left(\frac{p_0-|\mathbf{k}|}{p_0}\right), \label{p - k Slashed}
\end{equation}
it is possible to compute $\Gamma$ through \cite{Kiessig:2010pr},
	%
	\begin{equation}\label{gammafulldef}
		\begin{aligned}
			\Gamma_i &= -\frac{1}{2p_0} \text{Tr}[(\slashed{p} + m_i )\text{Im}\Sigma] = \frac{-(YY^\dagger)_{ii}}{4p_0}(2 \pi) \times \\ & \int \frac{d^3k}{(2\pi)^3} \text{Tr}\left[(\slashed{p}+m_i)(P_R) \frac{\slashed{p}-\slashed{k}}{4 \mathbf{|k|} \mathbf{|p-k|}}\right]  (1+n_B(\mathbf{|k|})-n_F(\mathbf{|p-k|}))\delta(p_0-\mathbf{|k|}-\mathbf{|p-k|}) \\
			&= \frac{-(YY^\dagger)_{ii} p^2}{16 \pi p_0^2 \mathbf{|p|}}\int^{k_+}_{k_-} dk (1+n_B(\mathbf{|k|})-n_F(\mathbf{|p-k|}))(p_0 - |\mathbf{k}|),
		\end{aligned}
	\end{equation}
where the angular integration was performed through the Dirac delta, which binds the moduli integration by imposing the appropriate bounds on the cosine function:
	\begin{equation}\label{deltatocos}
		\delta(p_0-|\mathbf{k}|-|\mathbf{p-k}|) = \frac{|\mathbf{p-k}|}{|\mathbf{p}||\mathbf{k}|} \delta(\cos\theta-\cos\theta_0), \\
	\end{equation}
	\begin{equation}
		\cos\theta_0 \equiv \frac{p_0}{\mathbf{|p|}}+\frac{\mathbf{|p|}^2-p_0^2}{2\mathbf{|p|}\mathbf{|k|}}, \quad
		k_\pm = \frac{1}{2}\left(p_0\pm |\mathbf{p}|\right).
	\end{equation}
Taking into account all contributions,
\begin{equation}
	\label{gammafullfinal}
	\Gamma_{N_i} = 4 \times \frac{1}{2}\frac{m_i^2}{\omega_{p_i}}\frac{{T^2(YY^\dagger)}_{ii}}{8 \pi |p| \omega_{p_i}} F(\mathbf{|p|} /T, m_i/T), 
\end{equation}
where $F(\mathbf{|p|}/T, m_i/T)$ is given by \cite{Bastero-Gil:2016qru}
\bq
\label{F(x,y)}
F(x,w) = f(x_+,\omega_p/T) - f(x_+,\omega_p/T) ,
\nq
with $x_{\pm} = k_{\pm}/T$, $\omega_p = \sqrt{\mathbf{p}^2 + m^2(T)}$ and
\begin{equation}
	\begin{aligned}
		f(x,w) =&  \int dx (w-x) \left[ 1 + \frac{1}{e^x-1}-\frac{1}{e^{(w-x)}+1} \right]  \\
		=& -\frac{\pi^2}{3}-w^2+xw-\frac{x^2}{2}+ (w-x)\ln\left(\frac{1-e^{-x}}{1+e^{-w+x}}\right) + \text{Li}_2\left(e^{-x}\right)+ \text{Li}_2\left(-e^{-w+x}\right).
	\end{aligned}
\end{equation}
where $w = \omega_p/T$. As a consistency check, it can be seen that taking the zero-temperature limit of \eqref{gammafullfinal} yields the result of \cite{delAguila:2005pin}.


\subsection*{Thermal Masses}

The thermal contribution to the masses of the right-handed neutrinos result from the real part of the fermion self-energy diagram
\begin{equation}\label{self-energy-eq}
	\Sigma_i = i \int \frac{d^4 k}{(2 \pi)^4} V_1 G^{(11)}(k) V_2 S^{(11)}(p-k),
\end{equation}
composed by a left-handed neutrino/Higgs loop. $V_{1,2}$ are the interaction vertices, $V_1=V_2 = P_L y_{ij}$, with a sum over the SM lepton families implied, where $\sum^3_{j=1} y^2_{1j} =\sum^3_{j=1} y^2_{2j} = y^2$.
Using the finite-temperature propagators
\begin{equation}
	\Delta(p) = \frac{i}{p^2-m^2+i\varepsilon} = i P\left(\frac{1}{p^2-m^2} \right) + \pi \delta(p^2-m^2), 
\end{equation}
where $P(x)$ is the Cauchy principal value. The real and imaginary parts of the propagators are \cite{Bellac:2011kqa}:
\begin{equation}
	\begin{aligned}
		G^{(11)}(k) &= \left[i P \left(\frac{1}{k^2} \right) + 2 \pi \delta (k^2) (1/2 + n_B(k))\right],\\
		S^{(11)}(k) &= \left[ i P \left(\frac{1}{k^2} \right) + 2 \pi \delta (k^2) (1/2 - n_F(k)) \right]\slashed{k},
	\end{aligned}
\end{equation}
for massless particles. Accounting for the factor $i$ in \eqref{self-energy-eq}, the real part of the self-energy comes from the imaginary component of the propagator product:
%
\begin{equation}
	\begin{aligned}
		&\text{Re}\Sigma_i (p)= \\&4P_Ly^2 \int \frac{d^4k}{(2 \pi )^4} \left[\frac{(\slashed{p}-\slashed{k})2 \pi \delta(k^2) (1/2 + n_B(k))}{(p-k)^2} +\frac{(\slashed{p}-\slashed{k})2 \pi \delta((p-k)^2) (1/2 - n_F(p-k))}{k^2}\right].
	\end{aligned}
\end{equation}
Redefining $k \rightarrow p-k$ in the second term, and separating the zero-temperature and finite-temperature contributions, $\Sigma = \Sigma^{0} + \Sigma^{T}$:
\begin{equation}
	\!\!\!\! \frac{\text{Re} \Sigma_i^{T} (p)}{4 y^2} =  P_L \int \frac{d^4k}{(2 \pi )^4} \left[(\slashed{p}-\slashed{k}) n_B(k) -\slashed{k}  n_F(k)\right] \frac{2 \pi \delta (k^2)}{(p-k)^2}.
\end{equation}
Unfolding $\delta(k^2) = \delta(k_0^2 - \mathbf{k}^2) = (\delta(k_0 \pm \mathbf{k}))/(2\mathbf{k})$, taking the nearly-massless limit ($p$ and $k$ are aligned), and performing the integration in $k_0$, one obtains
%
\begin{equation}
	\begin{aligned}
		&\text{Re} \Sigma_i^{T} (p) = \\& \frac{2P_L y^2}{4 \pi^3} \int \frac{d^3k}{2 k} \left[\frac{(p_0-k) n_B(k) -k  n_F(k)}{2 k (p-p_0) + (p_0^2-p^2)}+ \frac{(p_0-(-k) n_B(-k) -(-k)  n_F(-k)}{2 k (p + p_0) + (p_0^2-p^2)}\right]\frac{\slashed{p}}{p_0} .
	\end{aligned}
\end{equation}
In the limit $p^2 \approx p_0^2$ and making use of
\begin{equation}
	n_B(-x) = -1 - n_B(x), \qquad n_F(-x) = 1 - n_F(x),
\end{equation}
one obtains
\begin{equation}
	\begin{aligned}
		\text{Re}\Sigma_i^T(p) &= \frac{y^2 P_L}{8 \pi^2 p_0 p^2}\slashed{p}  \int dk \bigg[ (p+p_0)\big((p_0-k)n_B(k) - k n_F(k)\big) + \\ &+  (p-p_0) \big(-p_0(1+n_B(k)) -k(n_B(k)+n_F(k))\big) \bigg]
	\end{aligned}
\end{equation}
Following \cite{Weldon:1982bn}, the leading $T^2$ contributions are generated by would-be quadratically divergent terms if no thermal distribution were present to act as a $k \sim \mathcal{O}(T)$ cut-off in the integration.
\begin{equation}
	\begin{aligned}
		\text{Re} \Sigma_i^{T'} (p) &= \frac{P_L y^2}{8 \pi^2 p_0 p^2} \slashed{p} \int dk \bigg( (n_B(k)+n_F(k))(p_0+p)-(n_B(k)+n_F(k))(p_0-p) \bigg)k \\
		& = \frac{P_L y^2}{4 \pi^2} \frac{\slashed{p}}{p^2} \int dk \left( n_B(k)+n_F(k) \right)k,
	\end{aligned}
\end{equation}
using $p=p_0$. Since
\begin{equation}
	\int^\infty_0 dk k n_B(k) = \frac{\pi ^2 T^2}{6}, \qquad \int^\infty_0 dk k n_F(k) = \frac{\pi ^2 T^2}{12},
\end{equation}
one obtains
\begin{equation}
	\text{Re} \Sigma_i^{T'} (p) =  \frac{P_L y^2}{p^2} \frac{T^2}{16} \slashed{p}.
\end{equation}
Taking into account all contributions, $\text{Re} \Sigma_f^{T'} (p) =  4\text{Re} \Sigma_i^{T'} (p)$, and the thermal contribution to the fermion masses is \cite{Weldon:1982bn}
\begin{equation}
	M_T^2=\frac{\text{Tr}}{4}\left[\slashed{p}\text{Re} \Sigma_f^{T'} (p)\right]  =  \frac{4 y^2 T^2}{4 \times16} \text{Tr}\left[ \frac{1-\gamma_5}{2}\slashed{p}\slashed{p}\right] =  \frac{y^2 T^2}{8}.
\end{equation}
%

\section{Dissipation coefficient}\label{AppC:DissCoef}
The relevant Lagrangian for the computation of the dissipation coefficient has two vertices for each Majorana fermion:
\begin{equation}
-\mathcal{L}^{\phi}_{\text{Yuk}} = \frac{M_i}{2} \overline{\mathbf{N}}_i \mathbf{N}_i + \frac{V_i}{2}\overline{\mathbf{N}}_i \delta \phi \left(e^{i \delta} P_R  + e^{-i \delta} P_L\right) \mathbf{N}_i. \label{YukLagExp}
\end{equation}
To obtain the correct result, one needs to take into account that the canonically normalized inflaton field is $\varphi=\phi/\sqrt{2}$. Doing so, the dissipation coefficient for each neutrino is given by \cite{BasteroGil:2010pb}
\begin{equation}
\begin{aligned}
\Upsilon &= \frac{1}{4T} \looppf \text{Tr}[\rho_{\psi_i} \mathcal{V}\rho_{\psi_i} \mathcal{V}]n_F(p_0) (1-n_F(p_0)). \label{DissCoefi}
\end{aligned}
\end{equation}
where $\mathcal{V}$ is the term inside parenthesis in \eqref{YukLagExp} times $V_i$, defined in \eqref{Vertices and Phase}. Note that one still has to sum over both fermionic contributions ($i=1,2$). The extra factor of $1/2$ compared to \cite{BasteroGil:2010pb} is due to the symmetry factors of the Feynman diagrams (related to having identical, Majorana, particles running in the loops). Expanding the fermionic trace above
\begin{equation}
\begin{aligned}
V_i^2 e^{2 i \delta}\text{Tr}\left[(\slashed{p}+M_i)P_R(\slashed{p}+M_i)P_R\right]&= 2M_i^2 V_i^2 e^{2 i \delta}, \\
V_i^2 e^{-2 i \delta}\text{Tr}\left[(\slashed{p}+M_i)P_L(\slashed{p}+M_i)P_L\right]&= 2M_i^2 V_i^2 e^{-2 i \delta}, \\
2V_i^2 \text{Tr}\left[(\slashed{p}+M_i)P_R(\slashed{p}+M_i)P_L\right]&= 4 M_i^2 V_i^2 . 
\end{aligned}
\end{equation}
where we used that $p^2 = M_i^2$. Using
\begin{equation}
\rho_B^2 = \frac{\pi}{2 \omega_{p_i}^2 \Gamma_{N_i}} \left(\delta(p_0 - \omega_{p_i})+\delta(p_0 + \omega_{p_i})\right),
\end{equation}
the dissipation coefficient can be written as
\begin{equation}
\Upsilon_i = \frac{V_i^2}{T} \cos^2\delta  \looppt \frac{M_i^2}{\omega_{p_i}^2 \Gamma_{N_i}} n_F(p_0) (1-n_F(p_0)). \label{DissCoef-Int}
\end{equation}
Summing over the two $\mathbf{N}$ contributions,
\begin{equation}
\Upsilon = \Upsilon_1 + \Upsilon_2 = \frac{V_1^2 + V_2^2}{T} \cos^2\delta  \text{Int.}, 
\end{equation}
where we used $M_1 \simeq M_2$, a good approximation when the thermal contribution dominates the right-handed neutrino masses. The integral is then defined as
\begin{equation}
\text{Int.} = \looppt \frac{M_i^2}{\omega^2_{p} \Gamma_{N}} n_F(p_0) (1-n_F(p_0)).
\end{equation}
Accounting for \eqref{gammafullfinal}, the dissipation coefficient is:
\begin{equation}
\Upsilon_i =\frac{\pi V_i^2}{y^2 T^3}  \cos^2\delta \looppt |\mathbf{p}| \frac{n_F(p_0) (1-n_F(p_0))}{F(|\mathbf{p}|/T, M_i/T)}, 
\end{equation}
where $y^2$ is the effective Yukawa coupling defined in \eqref{LowTDissCoef}, with $F(|\mathbf{p}|/T, M_i/T)$ is defined in \eqref{F(x,y)}.
Performing the integration over the solid angle and changing the integration variable to $x\equiv |\mathbf{p}|/T$,
%
\begin{equation}
\label{DissCoef}
\Upsilon_i = \frac{\pi}{4 \pi^3} \frac{V_i^2}{y^2} \cos^2(\delta) T  \int_{0}^{\infty} \frac{d^3p}{T^3} \frac{|\mathbf{p}|}{T} \frac{n_F (1-n_F)}{F(|\mathbf{p}|/T, M_i/T)} =\frac{2V_i^2}{\pi y^2} T \cos^2(\delta) \int_{0}^{\infty} dx x^3 \frac{n_F (1-n_F)}{F(x, w)}. 
\end{equation}
where $w \equiv M_i/T$. In order to find a simple expression for the dissipation coefficient \eqref{DissCoef}, we analyzed two asymptotic regimes: i) the high-temperature limit ($w\ll1$) and ii) the low-temperature limit ($w\gg1$). The final result is an interpolation between the two. 

\begin{figure}[h]
	\centering\includegraphics[width=0.6\textwidth]{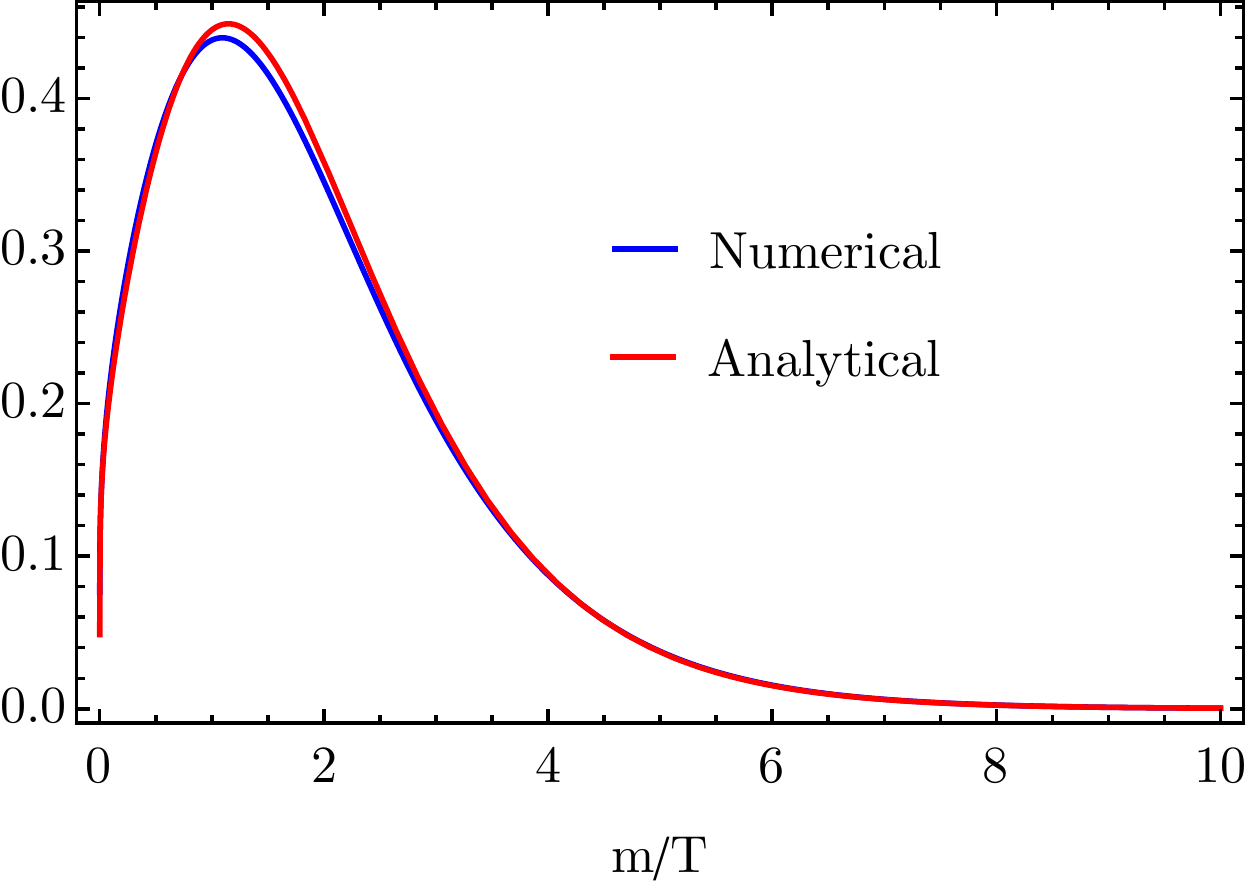} 
	\caption{Numerical evaluation of the integral in \eqref{DissCoef} and its analytical approximation, $\Upsilon_i/B$ with $B = T \cos^2(\delta)V_i^2/y^2$. \label{fig:NumericalDiss}}
\end{figure}

Assuming $x \sim 1$, which is the most relevant region of the integral, the high-temperature limit becomes
\begin{equation}
F(x, w) \sim -2x\log(w) \Rightarrow \int_0^\infty dx \frac{x^2}{\beta - 2\log(w)} n_F (1-n_F) = \frac{\pi^2}{6(\beta - 2\log(w))},
\end{equation}
where $\beta$ is a constant to be determined. Numerically, we find that $\beta=3$ results in a low error in the estimate of the dissipation coefficient in the high-temperature limit. For the low-temperature case,
\begin{equation}
\begin{aligned}
F(x,w) &\sim \frac{xw}{2}\coth{\left(\frac{w}{2}\right)} \Rightarrow \int_0^\infty dx \frac{x^2}{w \coth{\left(\frac{w}{2}\right)}} e^{-w-\frac{x^2}{2w}} \left(1-e^{-w-\frac{x^2}{2w}}\right) \\ &= \frac{1}{2} \sqrt{\pi}e^{-2w} \left(2 \sqrt{2} e^w-1\right)\sqrt{w}\tanh{\left(\frac{w}{2}\right)}
\end{aligned}
\end{equation}
Finally, we may introduce Boltzmann factors to combine these two asymptotic limits in a smooth way, yielding:
\begin{equation}\label{DissCoef-final}
\Upsilon_i=\frac{V_i^2}{y^2} T \cos^2\delta  \left( \frac{\pi e^{-2(M_i/T)}}{(9-6e^{-(M_i/T)}\log{(M_i/T)})} +  \frac{ e^{-(M_i/T)}(2\sqrt{2} - e^{-(M_i/T)})  \tanh{(M_i/T)}}{\sqrt{\pi}} \right), 
\end{equation}
which features a relative error never greater than $4\%$ (see Figure \ref{fig:NumericalDiss}).
Taking the thermal corrections to be the dominant contribution to the heavy neutrino masses during inflation, i.e., $M_t \gg M_0$, then $M_i \propto M_t T$ and the contributions from $N_1$ and $N_2$ are easily combined:
\begin{equation}\label{cosine}
\Upsilon =\frac{ \left( g_1^2 +g_2^2 \right)}{2 y^2} T  \cos^2(\delta)  \left(\frac{\pi e^{-M_t}}{9 e^{M_t}-6\ln{M_t}}+\frac{e^{-2M_t} \left(2\sqrt{2}e^{M_t}-1 \right)\sqrt{M_t}\tanh{(\frac{M_t}{2})}}{\sqrt{\pi}}\right) . 
\end{equation}
Replacing $M_T$ with the result obtained from the thermal mass computation, one arrives at
\begin{equation}\label{CT-r neq 0}
\Upsilon=\frac{ (g_1^2+g_2^2)}{2y^2} \cos^2\delta \left(\frac{\pi}{9-6\ln(y/2\sqrt{2})}\right) T \equiv C_T T, 
\end{equation}
which is a function of the three coupling constants relevant to inflationary dynamics: $g_1, g_2$, and $y$.


\end{document}